\documentclass[final,5p,times,twocolumn]{elsarticle}

\makeatletter
\def\ps@pprintTitle{%
  \let\@oddhead\@empty
  \let\@evenhead\@empty
  \def\@oddfoot{\reset@font\hfil\thepage\hfil}
  \let\@evenfoot\@oddfoot
}
\makeatother

\usepackage[inkscapelatex=false]{svg}
\usepackage[flushleft]{threeparttable}
\usepackage{verbatim}

\usepackage{graphics}
\usepackage{lscape}

\usepackage{tikz}
\newcommand*\emptycirc[1][1ex]{\tikz\draw (0,0) circle (#1);} 
\newcommand*\halfcirc[1][1ex]{%
  \begin{tikzpicture}
  \draw[fill] (0,0)-- (90:#1) arc (90:270:#1) -- cycle ;
  \draw (0,0) circle (#1);
  \end{tikzpicture}}
\newcommand*\fullcirc[1][1ex]{\tikz\fill (0,0) circle (#1);}

\usepackage{pifont}

\usepackage{multirow}
\usepackage{hyperref}
\usepackage{enumitem}
\usetikzlibrary{shapes.geometric, positioning}
\usepackage{float}
\usepackage{makecell}

\usepackage{amssymb}

\begin{document}

\begin{frontmatter}



\title{Generative AI for Internet of Things Security: Challenges and Opportunities}


\author[a]{Yan Lin Aung\corref{cor}\fnref{eqc}}
\cortext[cor]{Corresponding author.}
\ead{yan\_lin\_aung@ieee.org}
\fntext[eqc]{Both authors contributed equally and ordered alphabetically.}
\author[a]{Ivan Christian\fnref{eqc}}
\author[b]{Ye Dong}
\author[a]{Xiaodong Ye}
\author[a]{Sudipta Chattopadhyay}
\author[a]{Jianying Zhou}
\affiliation[a]{
  organization={Singapore University of Technology and Design},    
  country={Singapore}
}
\affiliation[b]{
  organization={National University of Singapore},  
  country={Singapore}
}

\begin{abstract}
As Generative AI (GenAI) continues to gain prominence and utility across various sectors, their integration into the realm of Internet of Things (IoT) security evolves rapidly.
This work delves into an examination of the state-of-the-art literature and practical applications on how GenAI could improve and be applied in the security landscape of IoT.
Our investigation aims to map the current state of GenAI implementation within IoT security, exploring their potential to fortify security measures further.
Through the compilation, synthesis, and analysis of the latest advancements in GenAI technologies applied to IoT, this paper not only introduces fresh insights into the field, but also lays the groundwork for future research directions.
It explains the prevailing challenges within IoT security, discusses the effectiveness of GenAI in addressing these issues, and identifies significant research gaps through MITRE Mitigations.
Accompanied with three case studies, we provide a comprehensive overview of the progress and future prospects of GenAI applications in IoT security.
This study serves as a foundational resource to improve IoT security through the innovative application of GenAI, thus contributing to the broader discourse on IoT security and technology integration.
\end{abstract}




\end{frontmatter}



\section{Introduction}
\label{sec:intro}
The development of Generative AI (GenAI) and Large Language Models (LLMs) signifies an advancement in artificial intelligence, distinguished by its capability to generate diverse content, including texts, images, and code.
This capability has brought GenAI tools into the spotlight, facilitating their integration into daily life to address various pertinent issues and tasks.
Given their utility in tasks such as data analysis and content generation, GenAI and LLMs are actively explored for their potential in more complex applications. 
GenAI has garnered significant attention in the field of cyber security, with recent studies underscoring its potential to enhance security measures, simulate attacks for training and testing, and refine threat detection systems through advanced data analytics.
Examples include studies conducted by~\citet{9105926},~\citet{10198233}, and~\citet{hassanin2024comprehensiveoverviewlargelanguage}, all of which focus on the potential advantages that GenAI could offer as a tool to automate various complex security tasks.

Internet of Things (IoT) is increasingly recognized as a critical area requiring detailed attention and innovative approaches, as IoT devices become more integrated into daily life and industrial systems.
As IoT devices are heterogeneous in nature, the security of these devices requires specialized knowledge and expertise.
GenAI has the potential to enhance existing methods or develop new approaches for IoT security, thereby reducing the need for specialist knowledge to implement advanced security solutions.
Consequently, GenAI represents a promising tool for future IoT security research, which could improve both the security and usability of IoT systems.
In the coming years, significant research is anticipated to be conducted on the use of GenAI to improve IoT security.

This paper presents a comprehensive survey of current state-of-the-art work on the application of GenAI to IoT security.
We begin by providing a foundational understanding of IoT systems along with the core principles of Generative AI (GenAI), with a primary focus on LLMs.
As part of our analysis of the current use of GenAI in enhancing IoT security, we explore the application of each model.
Subsequently, we analyze potential further applications, identifying areas where GenAI could be beneficial with three case studies.
Our evaluation is articulated through the use of the MITRE ATT\&CK Mitigation framework for Industrial Control Systems (ICS).
\subsection{Internet of Things (IoT)}
IoT is a transformative concept in connectivity, where an extensive network enables devices ranging from household appliances to medical equipment to connect directly to the Internet, facilitating seamless data exchange without human intervention.
This innovation has broad applications in smart homes, healthcare, transportation and urban development, significantly improving operational efficiency~\citet{kimani2019cyber}.
~\citet{alwahedi2024machine, chui2023survey} present a similar perspective, who describe the IoT as a network that connects physical objects through embedded sensors and software.
This configuration not only facilitates the exchange of real-time data, but also transforms physical data into digital information, providing a comprehensive means of managing diverse systems.
The framework underscores the ability of the IoT to digitalize physical entities, fostering an intelligent and interconnected environment.

As described in a study by~\citet{hassija2019survey}, the IoT ecosystem consists of four essential layers: the foundational layer that uses sensors and actuators for data collection, followed by a communication layer that transmits the data.
The middleware layer then bridges the data flow between the network and application layers, allowing processing and integration.
The final layer hosts various IoT applications, such as smart grids and smart factories, demonstrating the structured and integrated approach of IoT systems in various sectors.
\subsection{IoT Security Challenges}
The potential vulnerabilities of IoT devices pose a significant security threat to IoT ecosystems.
Their interconnected nature exposes them to a variety of cyber threats, data breaches, and privacy violations~\citet{hassija2019survey}.
Insufficient security updates, inadequate security measures, and difficulties in managing dynamic device configurations are among the most common security issues. 
These vulnerabilities mainly consist of communication vulnerabilities, operating system vulnerabilities, and software vulnerabilities. 
There is an ongoing research effort to enhance the overall security of IoT to effectively address these issues.

The diverse application of IoT devices, from home automation to medical systems, makes them an attractive target for malicious activity.
Therefore, it is imperative to implement protective measures such as authentication protocols, intrusion detection systems, and machine learning algorithms to fortify these networks against potential threats.
Various methods have been employed to address these issues, including deep learning~\citet{9060970} and blockchain technology~\citet{10.1145/3320154.3320163}.
\subsection{Generative AI and Large Language Models}
In the context of GenAI development, LLMs could be viewed as a breakthrough in AI innovation due to their ability to generate, classify and reason based on the datasets with which they are trained~\citet{jo2023promise}.
Through advancement in algorithms and computational power, GenAI has become increasingly important in a wide range of domains, including cyber security.
With the capability to generate novel data instances from learned patterns, this technology offers a revolutionary approach to data analysis and simulation, demonstrating its potential for transformative applications in the digital world.
LLMs, such as ChatGPT~\citet{openai2024gpt4} and Gemini~\citet{10113601}, represent a significant advancement in GenAI, particularly in the processing and generation of natural language.
These models have evolved to understand context, generate coherent responses, and even detect anomalies in text, making them invaluable tools that extend beyond simple communication.
LLMs demonstrate the increasing sophistication of AI's ability to handle complex, nuanced tasks, mirroring human-like understanding and interaction with large volumes of data.

Digital defense strategies have been transformed by the integration of GenAI, specifically LLMs, into cyber security.
LLMs are well-positioned to enhance security for interconnected digital systems, including IoT.
The use of these technologies for security purposes is gaining traction, indicating a promising direction for addressing security threats.
The emerging field of GenAI, particularly through the lens of LLMs, represents not only a technological advancement but also a transformative force in cyber security.
These AI models generate realistic simulations and learn complex patterns providing significant benefits.
\section{Background} \label{background}
In recent years, cyber threats have highlighted the importance of developing adaptive defense mechanisms against evolving attack vectors to prevent situations such as a botnet attack by advanced persistent threat (APT)~\citet{Daws_2024}.
As the cyber security landscape evolves, GenAI has emerged as an essential tool for enhancing security measures.
GenAI, characterized by its ability to produce new content that mimics real-world phenomena, facilitates a spectrum of security applications, from passive threat detection to active mitigation.

In the realm of security, the evolution of GenAI has underscored its role as a double-edged sword.
On the one hand, these technological advances represent a significant step forward in digital transformation, enhancing security through automated responses, threat intelligence, and malware detection~\citet{10198233}.
Their ability to generate highly realistic content across various mediums illustrates the potential to increase threat detection and response measures.
On the other hand, the same capabilities that contribute to security enhancements also present new vulnerabilities, as GenAI models have been exploited by rogue actors for offensive purposes.
This includes generating deepfake videos for disinformation campaigns, crafting convincing phishing emails, and spreading misinformation on social media, introducing new challenges and risks~\citet{eze2024analysis,mitra2024world}.
\subsection{Evolution of AI in Security}
In this section, we briefly review the landscape of AI within cyber security research. 
\subsubsection{Machine Learning (ML)}
Detection methods in network security initially relied on traditional machine learning algorithms in the early days.
These methods process large volumes of log data, identify specific patterns, and perform verification.
Techniques such as linear regression and decision trees effectively handle massive data and work well in practical applications.
For example, some machine learning-based intrusion detection systems (IDS) analyze user or device behavior to identify abnormal patterns.
Security operation centers (SOC) use machine learning to detect abnormal activities that deviate from normal behavior.
These methods enhance efficiency and flexibility in security monitoring, enabling real-time threat detection.
However, the analysis of patterns and anomalies is static and requires retraining and tuning should the threats change, especially in the field of cyber security~\citet{zeadally2020harnessing}.
\subsubsection{Deep Learning (DL)}
With larger and more complex behaviors being collected, the limitations of traditional ML methods become apparent in terms of the models' capabilities to predict, classify, and learn effectively.
There arises a need to solve more robust, comprehensive, and large-scale problems that ML algorithms would have difficulty solving.
The evolution of ML is followed by the development of deep learning (DL) algorithms, which address more complex problems such as image recognition~\citet{Alzubaidi2021-dk}, sentiment analysis~\citet{zhang2018deep}, deep anomaly detection~\citet{pang2021deep}, and natural language processing~\citet{ghosh2016contextual}.
The ability of deep learning to solve more complex problems has become a foundation for further improvements.
\subsubsection{Generative Adversarial Networks (GANs)}
Following the adoption of DL, Generative Adversarial Networks (GANs) have introduced a novel dimension to cyber security.
The application of GANs is twofold: enhancing security defenses by increasing their ability to detect sophisticated threats~\citet{park2022enhanced, yinka2020review}, and contributing to the development of complex threats such as AI-driven malware or phishing emails.
A growing number of attacks are leveraging AI-driven techniques as threat actors evolve their strategies. This approach, when combined with conventional attack methods, enables attackers to inflict even greater damage~\citet{kaloudi2020ai}.
\subsubsection{Recent Application of GenAI}
Generative Pre-trained Transformers (GPTs) represent the latest advancement in this evolution, extending the capability of AI into natural language processing.
This development has a notable contribution to security, as it could be used to enhance security, for example, by developing robust security policies to protect against ransomware attacks.
A study comparing GPTs with conventional policy-making sources found that GPT-generated policies outperform those derived from security vendors and government agencies in terms of effectiveness and ethical compliance, particularly with tailored input and expert oversight~\citet{mcintosh2023harnessing}.
GPTs could also be used to investigate the potential for AI misuse~\citet{renaud2023chatgpt}, such as generating malware using LLMs~\citet{pa2023attacker, greshake2023not}.
Indirect prompt injection facilitates remote exploitation of LLM-integrated applications, posing threats such as data theft and contamination of information ecosystems.
Several practical demonstrations emphasize the risks associated with the execution of arbitrary code and the manipulation of functionality.
\subsection{Applications of GenAI in Security}
GenAI has significantly transformed security practices by introducing advanced capabilities for threat detection, simulation, and data protection.
The applications of GenAI in this domain include, but are not limited to, the following:

\vspace{1em}
\noindent \textbf{Enhanced Threat Intelligence:} The GenAI model is able to analyze large amounts of data to predict and simulate emerging threats, providing security professionals with information on potential vulnerabilities and attack vectors~\citet{gupta2023chatgpt, alwahedi2024machine}.
Organizations must understand the characteristics of new and evolving threats to prepare more effectively and ensure that they remain one step ahead of cybercriminals.

\vspace{1em}
\noindent \textbf{Sophisticated Phishing Attack Simulations:} GenAI assists in the development of more effective training programs due to its ability to generate convincing phishing emails and social engineering tactics~\citet{bethany2024large}.
Employees are educated about the nuances of phishing attacks through these simulations, thereby significantly reducing the likelihood of successful breaches.

\vspace{1em}
\noindent \textbf{Automated Security Testing:} GenAI could automate the creation of test cases for secure software, ensuring that applications are robust against a wide range of attacks~\citet{hilario2024generative, deng2023pentestgpt}.
This involves generating malicious inputs to test the resilience of systems to injection attacks and other vulnerabilities.
This capability is crucial in sectors such as banking and e-commerce, where identity theft poses significant risks.

\vspace{1em}
\noindent \textbf{Synthetic Identity Fraud Detection:} GenAI models could help in designing algorithms that detect and prevent fraudulent activities by understanding patterns of synthetic identity fraud~\citet{ahmadi2023open}.

\vspace{1em}
\noindent\textbf{Adaptive Defense Mechanisms:} GenAI models are capable of simulating a variety of attack scenarios, enabling security systems to create dynamic defensive strategies~\citet{neupane2023impacts, kucharavy2023fundamentals}.
This approach helps to develop resilient systems that could defend against sophisticated and adaptive threats.
~\citet{sai2024generative} describe ten security products that leverage GenAI to enhance their security measures.
These include Google Cloud Security AI Workbench, Microsoft Security Copilot, and SentinelOne Purple AI.
Additionally, 11 applications of GenAI were identified in the security domain, including threat intelligence, security questionnaires, bridging the gap between technical experts and non-experts, vulnerability scanning and filtering, and secure code generation.
\subsection{Applications of GenAI in IoT Security}
As researchers explore applications and investigations involving GenAI, we observe early works, though emerging, on the use of GenAI in IoT security given the increasing prevalence of IoT devices and their vulnerabilities.
Therefore, further investigation is necessary on how GenAI could be integrated to improve IoT security measures and strategies.
Relevant publications have been gathered to compile and explore this area, examining how GenAI could address IoT security.
Our findings highlight useful ideas and set the stage for future research in this area.
\section{Survey Methodology}
\label{sec:search-method}
Our research methodology focuses on gathering papers from conferences, journals, workshops, and publications centered on or related to our investigation on the application of GenAI for IoT security. Our search process includes several different search methods, databases, and search engines to ensure that we collect as much relevant work as possible.
\subsection{Search Methods}
We used a structured approach in our literature search to have a comprehensive coverage and relevance of our survey on the intersection of GenAI and IoT security.

\smallskip
\textbf{OWASP Framework Insights:} Our search strategy was enrched by the inclusion of keywords from the OWASP IoT Top 10 and GenAI-related terms, such as ``weak passwords + IoT + Large Language Model''.
This approach uncovers research addressing the security challenges identified by OWASP and areas yet to be explored by GenAI solutions.
By analyzing the findings from these searches, we evaluated the potential role that GenAI could play in enhancing IoT security.

\smallskip
\textbf{MITRE ATT\&CK Framework Integration:} We also incorporated the MITRE ATT\&CK framework, focusing on tactics and techniques pertinent to Industrial Control Systems (ICS).
The application of ICS matrix from this framework provided a structured approach for identifying and analyzing threats specific to the Industrial Internet of Things (IIoT) sector.
Keywords such as ``hardcoded credentials + IoT + LLMs'' were employed to refine our search, ensuring a focused examination of the literature.
\subsection{Sources}
As part of the compilation of our sources for the application of GenAI in IoT security, we systematically gathered research from various conferences and journals, enhanced by contributions from leading academics and expanded searches in public repositories.
During the selection process, we sought to include works that have made significant contributions to cyber security and IoT, which have been subjected to rigorous peer reviews and are relevant to our study.

\smallskip
\textbf{Academic Publications: } We prioritized the sourcing of conferences and journals well known for their contributions to cyber security and IoT.
This included key venues such as IEEE Transactions on Dependable and Secure Computing (TDSC), Transactions on Information Forensics and Security (TIFS), International Journal of Critical Infrastructure Protection (IJCIP), Transactions on the Internet of Things (TIOT), IoT Journal, Computers \& Security, and ACM Transactions on Privacy and Security (TOPS).
Each source was selected for its relevance, rigorous peer review, and ability to provide the most recent and impactful research findings.

\smallskip
\textbf{Other Sources: } Considering all the recent research on GenAI, there are likely many studies yet to be accounted for in publications.
As such, papers available on arXiv are considered as a possible source of contributions if the papers have relevant implementations and results.
We have also looked into possible GitHub repositories to account for related works.
This allows us to include works that have not yet been published as part of the possible application of GanAI in improving IoT security.
\section{Source Analysis Using MITRE ATT\&CK ICS Mitigation Techniques}
\label{sec:detection}
\begin{figure*}[!ht]
    \centering
    \includegraphics[width=0.95\textwidth,keepaspectratio]{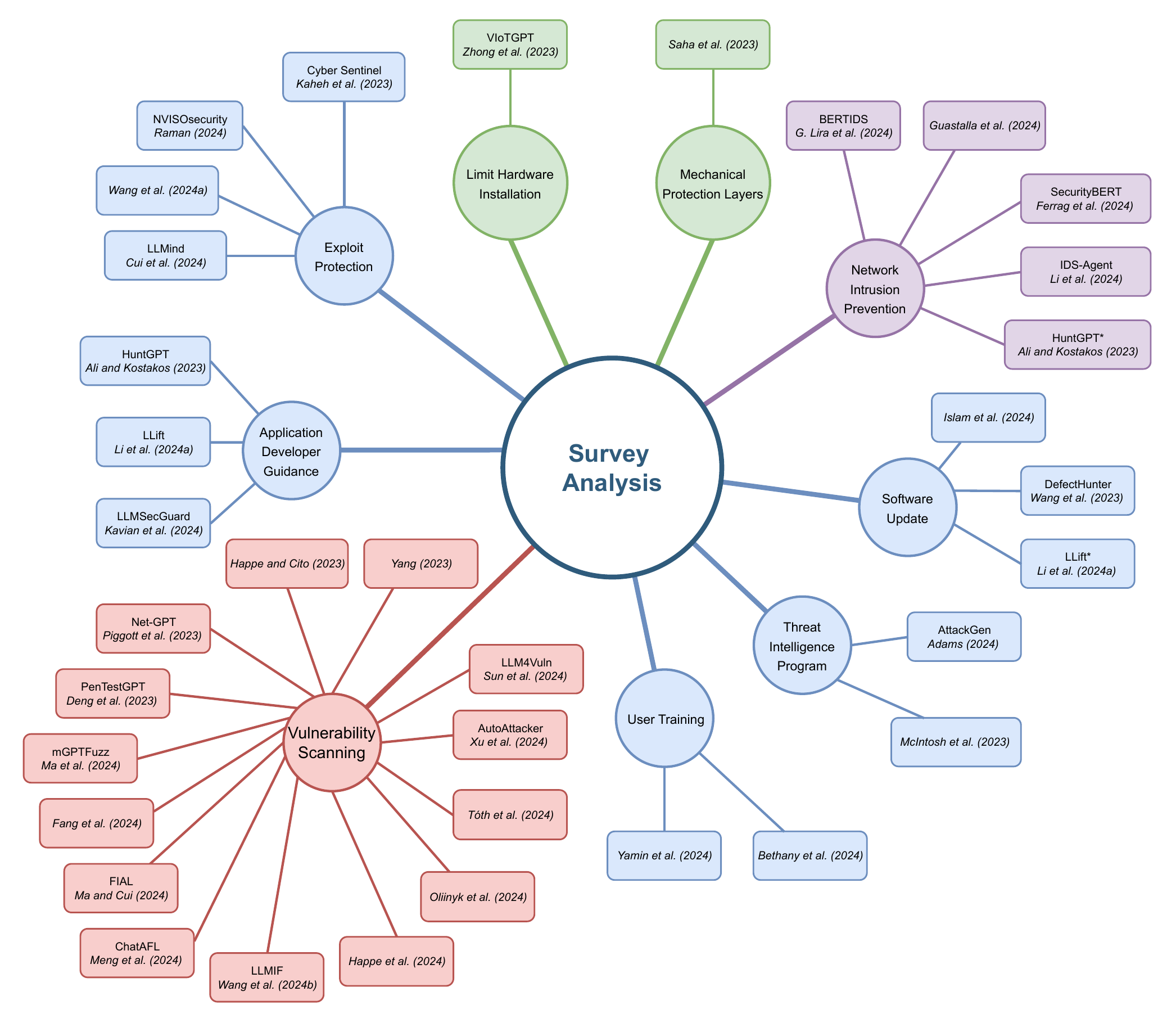}
    \caption{Mapping GenAI Applications for Cyber Security to MITRE ATT\&CK ICS Mitigations Framework}
    \label{fig:genai-attack-mappings}
\end{figure*}

We rely on ATT\&CK ICS Mitigations framework~\citet{icsmitre} to categorize the techniques used in the sources as illustrated in Figure~\ref{fig:genai-attack-mappings}.
IoT systems have similar, if not all, methods to compromise them as ICS.
Although sources may not directly address the issue in the context of IoT, they are included due to their relevance.

In the following subsections, we first explain the details and findings about the mitigation techniques.
It should be noted that there are some overlaps, as a study might address more than one mitigation technique.
Subsequently, we provide the potential for using GenAI to secure IoT networks and systems.
Each work is discussed according to the following capabilities:
\begin{itemize}
\item \textbf{External Threat Detection (ETD)} refers to the ability of GenAI to detect or prevent external threats, be malicious or aimed at scanning vulnerabilities (e.g., fuzzing).

\item \textbf{Internal Anomaly Detection (IAD)} measures GenAI's ability to identify and detect anomalies within the system, such as secure coding practices, vulnerabilities in software or hardware components.

\item \textbf{Response Automation (RA)} measures the ability of LLMs to generate an automated response based on their functions.
These responses could take the form of alerts, documents, or software patches, ultimately improving the security of the IoT system.

\item \textbf{Research Maturity (RM)} measured by the security field that GenAI addresses and potentially improves.
Each GenAI focuses on a specific security niche and aims to resolve existing issues within that niche.
If the field is not well explored, it indicates low research maturity and is ripe for exploration and development.
In contrast, if established standards or tools are available to improve system security, the niche exhibits high maturity.

\item \textbf{Development Potential (DP)} measures GenAI's potential for further development of the current implementation evaluated based on the tool's ability to address specific security functions.
In particular, GenAI application with high DP implies that the application itself has the potential to be scaled or extended towards other security function while low DP means the application is self-contained and rather complete.

\item \textbf{Impact on Security (IS)} refers to the significance and scope that the proposed GenAI tool is capable of addressing in the security field.
\end{itemize}
It is important to note that certain capabilities are mutually exclusive.
For instance, a particular GenAI application may exclusively focus on ETD without addressing IAD, and vice versa.
We reviewed 33 state-of-the-art works and provided our analysis on the potentials and impact of application of GenAI in IoT security.
In Table~\ref{tab:potentials-genai-iot-sec}, a full circle indicates that the proposed GenAI application is comprehensive for one of the six capabilities, whereas a half circle means that the GenAI application addresses the capability with certain limitations.
Lastly, an empty circle indicates that the GenAI application does not cover a particular capability.
For each work, we provided justifications for the evaluation of each of the six capabilities in Table~\ref{tab:potentials-details} without being too text-heavy.
\subsection{Application Developer Guidance} \label{subsec:adg}
This section discusses the use of GenAI to guide software developers in creating secure software from the outset.
The tool helps developers by providing guidance on secure software development or preparing development policies for software security.
Since most IoT devices and systems consist of a significant portion of software, this would benefit developers of IoT devices and systems.

\smallskip
\noindent\textbf{LLMSecGuard~\citet{Kavian2024LLMSG}} focuses on minimizing vulnerabilities and hard-coded credentials in production code.
Through the use of a static code analyzer, LLMSecGuard iteratively analyzes the code to identify vulnerabilities.
The code and analysis results are then presented to developers to create secure software that demonstrates the capabilities of LLMs.
A fine-tuned LLMSecGuard could be used to assist in secure software development for IoT systems.
For example, incorporating LLMSecGuard as part of the testing phase during software development ensures code security for software in the IoT.
This enables developers to create secure software without hard-coded credentials or keys oversight.
The impact of LLMSecGuard on the security field is limited to automating the patching process.

\smallskip
\noindent \textbf{LLift~\citet{libugdetection2024}} investigated Use Before Initialization (UBI) variables within the Linux kernel to uncover any bugs.
LLift has a 50\% precision rate, with 5 out of 10 reported positives being true vulnerabilities.
It also has a 100\% recall rate, as it did not misidentify any real bugs in the Rnd-300 dataset~\citet{li_2024_10780591}.
LLift helps users create secure code by identifying UBI bugs.
LLift could be implemented for IoT systems running embedded Linux, effectively identifying UBI bugs.
By fine-tuning LLift, developers could remove UBI bugs during testing and production, enhancing coding security, and mitigating risks in IoT systems.

The development potential for LLift includes adding an automated patcher and improving explainability and live detection.
UBI bugs are critical in Linux kernels, potentially leading to privilege escalation and information leakage.
LLift is among the few tools that address UBI bugs and its role in detecting these bugs is crucial for IoT security, especially given the use of embedded Linux by various vendors.
With further improvements, LLift could significantly impact IoT security by effectively patching UBI bugs.

\smallskip
\noindent\textbf{HuntGPT~\citet{ali2023huntgpt} : } GenAI, LLMs in particular, remains a black box in terms of its training and decision-making process.
To ensure that LLMs provide alerts with minimal false positives and are understandable by experts, it is necessary to explain why LLMs generated the alerts.
HuntGPT was developed for this purpose, trained in the KDD99 dataset~\citet{kdd99dataset} as an anomaly detector for benchmark attacks and standardized cyber security certification exams.
HuntGPT has a success rate of more than 70\% in these exams, demonstrating its knowledge and understanding.
It includes a dashboard that explains the attacks in the dataset, creating an explainable AI for cyber security experts to assess the reasoning behind the generated alerts.
Using IoT-related datasets such as NSL-KDD~\citet{nslkdd2009} or IoT attack benchmarks~\citet{iotattackbenchmark} could make attacks and anomalies explainable for IoT with HuntGPT, possibly with fine-tuning.
This helps experts quickly understand problems in IoT systems.
Implementing LLMs to generate a dashboard explaining anomalies could definitely be applied to IoT.

HuntGPT is highly automated and capable of detailed analysis and recommendations.
It has advanced automated response capabilities for IoT security, a field that is still emerging and integrating with other systems.
More IoT datasets could further fine-tune the tool for specific applications.
HuntGPT has significant potential for development into a live IDS, capable of addressing both external and internal threats.
In general, HuntGPT significantly impacts security by automating the analysis and classification of data sources, streamlining the process of identifying and understanding potential threats.
\subsection{Exploit Protection}
This section describes the application of GenAI to protect systems from exploits.
Protection could be achieved by blocking code execution and automated scripts.
LLMs could highlight important components and automate the hardening of system security in IoT environments.

\smallskip
\noindent \textbf{LLMind~\citet{cui2024llmind}} is an assistant that could perform complex tasks on an IoT network, acting as a gateway to control various devices.
It is used to control a Wi-Fi router, a mobile robot, and security cameras in a smart home system.
Implementation of LLMind is similar to giving prompts to a secretary, who then executes tasks based on generated finite-state machine code.
The study shows that LLMind successfully performed tasks such as object detection, human recognition, and report generation.
Although not directly related to security, LLMind has the potential to protect systems from exploits by executing device-specific security hardening techniques.
For example, it could update the allowed IP list for a security camera if an unauthorized remote connection is detected.
This potential of LLMind to receive prompts, generate scripts, and execute them could be exploited for the security of IoT.
LLMind is able to complete tasks autonomously for the given queries and could be improved to execute security-specific tasks.
The tool's current impact on security is limited due to its specific capabilities in physical and network protection.

\smallskip
\noindent \textbf{\citet{wang2024hybrid}} experimented with preventing attackers from escalating privileges by creating an LLM to identify user privilege-related variables (UPR) through fine-tuning with specific UPR knowledge.
The LLM identified the UPR with a 13.49\% false positive rate in typical programs where the UPR score was more than 80\%.
This LLM helps security analysts prioritize security enhancements for UPR variables, potentially creating more secure systems.
In IoT systems, it could be implemented in devices or network edges to secure critical code, such as OS-level programs.
With fine-tuning and IoT domain knowledge, the LLM could assist in the identification of UPR and help prevent privilege escalation, making it a valuable tool for mitigating attacks from exploits.
In terms of response automation, this LLM identifies UPR variables and prompts the user with minimal human intervention.
Extensive research on security of these critical variables leaves little room for further improvements.
However, this implementation could be a pioneering tool for identifying critical variables within IoT systems.
The security impact of the tool is significant as it could identify vulnerable UPR variables.

\smallskip
\noindent \textbf{NVISOsecurity~\citet{Raman_2024}} is an advanced LLM tool designed to protect vulnerabilities using an adversary emulation platform called Caldera.
Developed with Microsoft's AutoGen~\citet{wu2023autogen} framework, it employs two LLMs to automate tasks in predefined scenarios, such as generating vulnerability reports or adversary profiles.
Although not directly implemented in IoT security, its customizability allows exploit protection.
Caldera plugins, such as Caldera OT, could be added to address MITRE ICS techniques relevant to IoT.
NVISOsecurity executes tasks via the terminal or PowerShell, altering the machine's state.
It blocks or terminates anomalous processes, demonstrating its potential to mitigate attacks through exploit protection.
NVISOSecurity requires minimal human interaction, generates automated responses, and executes commands to prevent external attacks.
The research maturity is growing with ongoing research in MITRE ATT\&CK and automated execution.
Its impact on security is significant, as it automates attack-defend simulations, streamlines the defense process, and allows security personnel to focus on other tasks.

\smallskip
\noindent \textbf{Cyber Sentinel~\citet{kaheh2023cyber}} was developed to create an LLM that could explain its actions (Explainable AI) and perform tasks to improve system security (Actionable AI).
It processes conversational queries to generate actions for security tasks.
In the study, Cyber Sentinel successfully analyzed user prompts to retrieve and block IP addresses connected to a machine within the last three hours.
Although simple for humans, this task requires multiple steps.
The results showed improved threat detection, operational efficiency, and real-time collaboration.
This demonstrates the potential of LLMs such as Cyber Sentinel for IoT.
With more domain-specific training, Cyber Sentinel could perform IoT-specific tasks, such as blocking of IP addresses or automated updates, to secure networks, highlighting the potential of LLMs in securing IoT systems.

Cyber Sentinel prevents external exploitation, demonstrating its ability to detect and act on external threats.
It automates security tasks based on user queries and requires minimal human input.
In terms of research maturity, Cyber Sentinel is part of an emerging field focused on automated task execution based on user queries.
The development potential includes adding more task executions and improving its capabilities for IoT systems.
Future research could also explore its use in IoT penetration testing.
The security impact of Cyber Sentinel is significant, enabling automation for both blue and red teams.
\subsection{Limit Hardware Installation}
LLMs could be utilized to restrict additional installations in an IoT system by employing an observer, such as a CCTV, to ensure the integrity of the physical installation and prevent unauthorized USB devices from being inserted.
Although not all studies directly address this mitigation technique for IoT, this section explores its potential use based on existing research.

\smallskip
\noindent \textbf{VIoTGPT~\citet{zhong2023viotgpt} : } To limit the installations of rogue devices, the traditional method uses CCTV to monitor the system.
However, this lacks intelligent alerts to detect anomalies.
VIoTGPT combines LLM with a vision-based model to handle tasks involving images and text queries.
It uses tools for face recognition, vehicle re-identification, anomaly detection, and action recognition, all fine-tuned with specific domain knowledge.
The output includes decisions, recommended tools, and tool output descriptions.
Fine-tuned with public video datasets, web-scraped data, and self-made datasets, VIoTGPT identifies and describes tasks such as anomaly detection and action analysis with 30-50\% accuracy in the test set and 60-70\% accuracy in the validation set.
By integrating LLM and image-based models, VIoTGPT could create descriptions and recommend tools for certain tasks.

For IoT systems, VIoTGPT's anomaly detection could mitigate insider threats by identifying actions like inserting rogue devices.
This allows VIoTGPT to alert users of potential threats and limit hardware installations, preventing rogue devices with malicious programs from being connected to the IoT system.
VIoTGPT requires human prompts and automatically provides visual and analysis results.
Possible improvements include automated task execution to prevent suspicious activities and other enhancements such as sound and speech analysis, active physical defense, and preventive actions on open ports in a physical IoT system.
However, its current implementation is limited to alerting and analyzing actions within an image.
\subsection{Mechanical Protection Layers}
This section discusses the application of LLM to enhance the protection of the mechanical layer in IoT, pertaining to the hardware of IoT devices.
This encompasses the security of the design and physical safeguarding of IoT devices exploring the potential of LLM to aid in designing secure hardware.

\smallskip
\noindent \textbf{\citet{saha2023llm} : } A critical aspect of IoT systems is hardware design, which could inherently contain vulnerabilities.
A vulnerable design could be exploited at the hardware level, making it difficult for software to prevent access by malicious actors.
In their study, Saha et al. trained an LLM to critique the design of system-on-chip (SoC) integrated circuits to evaluate their security.
Although not specifically focused on IoT, this study demonstrates the potential of LLM to assess the security of SoC design.
Tests such as security verification, countermeasure development, security assessment, and vulnerability insertion were conducted to create a more secure SoC.
These tests are crucial because SoC security depends on a human-mistake-free and vulnerability-free initial design.
LLM's adaptability allows for dynamic implementation of security tasks in SoC design.
The study also suggests that LLM could improve the security of current and future SoC designs, helping to patch hardware design vulnerabilities in IoT devices and systems.

The proposed tool addresses T0880 tactic and prevents vulnerabilities and exploits from hardware design.
Its response automation capabilities depend heavily on user instructions and security rules.
In terms of research maturity, very few works address hardware design to improve security.
The development potential includes full automation of design critique to minimize user inputs and support diverse security standards.
Design critique and improvement could defend against zero-day vulnerabilities from hardware weaknesses and prevent hardware exploits and side-channel attacks.
This is the only LLM implementation that addresses possible exploits in an IoT setting using Mechanical Protection Layers.
\subsection{Network Intrusion Prevention}
This section explains how LLM protects IoT systems from network intrusions in ICS by using network intrusion detection or prevention modules.
Although not all studies focus on IoT, their potential for IoT implementation is discussed.

\smallskip
\noindent \textbf{BERTIDS~\citet{lira2024}} is a LLM-based tool for network intrusion detection.
It processes and understands network log data to identify and classify anomalies.
BERTIDS is adaptive and continuously learns new behavior to combat new threats, allowing it to detect network attacks that evade rule-based detectors.
Using the NSL-KDD dataset~\citet{nslkdd2009}, BERTIDS achieved the highest accuracy, precision, and F1 score (above 98\%) compared to other methods.
Although not directly implemented in IoT datasets, BERTIDS could be adapted to IoT by using the IoT attack benchmark dataset to identify attacks such as DoS, web-based, and Mirai.
As IoT evolves, LLMs such as BERTIDS could be adapted to understand network communication patterns.
BERTIDS shows significant potential for developing LLM implementations in detecting network intrusions.

\smallskip
\noindent \textbf{\citet{guastalla2023application}} conducted a study to detect DDoS attacks using LLM.
They trained and fine-tuned an LLM with CICIDS2017 and Urban IoT datasets to identify DDoS attacks.
The results demonstrated that the LLM achieved more than 90\% accuracy for both datasets when trained with few-shot learning methods.
However, the study has not been tested in a real network setting, which could impact its accuracy.
Despite this, the research shows potential for using and improving LLMs to detect network intrusion anomalies, such as DDoS attacks, within IoT systems.
The proposed tool is automated and requires minimal human intervention.
Its development potential lies in the ability to detect different types of attacks.
In terms of its impact on security, this was among the early works using LLM as an anomaly detector or IDS.

\smallskip
\noindent \textbf{SecurityBERT~\citet{ferrag2024revolutionizing}} utilizes BERT to create a lightweight model for IoT.
The study used network data to generate anomaly detection within the system, focusing on DDoS, information gathering, malware, injection, and man-in-the-middle attacks.
SecurityBERT outperformed traditional ML and DL techniques with 98.2\% accuracy, while other techniques were around or below 97\%.
It was integrated into a real-life setting, using internal network traffic within the IoT system.
This implementation demonstrates that SecurityBERT is a successful anomaly detector to identify different types of attack within an IoT system.
SecurityBERT is able to classify external threats based on network traffic features and autonomously generates classification results.
It is already trained with an IoT-related dataset.
A possible research direction for SecurityBERT is to create an agent that acts on the classification results.

\smallskip
\noindent \textbf{IDS-Agent~\citet{li2024idsagent}} is a very recent work on the intrusion detection system for IoT, using LLM to improve detection.
Unlike traditional IDS methods, it combines reasoning and action for better performance and zero-day attack detection.
In experiments, IDS-Agent outperformed state-of-the-art machine learning-based IDS and previous LLM-based methods, achieving F1 scores of 0.97 on the ACI-IoT benchmark and 0.75 on the CIC-IoT benchmark.
IDS-Agent is able to detect zero-day attacks with a recall of 0.61 surpassing previous approaches specially designed for this task.
The IDS-Agent automatically detects and classifies attacks targeting the IoT.
Though it was validated using two datasets, it could be further developed and extended towards live detection and become a more impactful security tool for first line of defence.

\smallskip
\noindent \textbf{HuntGPT~\citet{ali2023huntgpt}} serves as an anomaly detector to create an Explainable AI for users.
As described in Section~\ref{subsec:adg}, it detects attacks in the dataset and displays them on a dashboard for user understanding.
The attack details help explain and understand the context.
Identifying attacks and anomalies, HuntGPT improves security, allowing users to address these issues.
This implementation of HuntGPT functions as both an anomaly detector and a tool for Explainable AI.
\subsection{Software Update}
This section outlines studies and experiments aimed at mitigating attacks and enhancing security by patching vulnerabilities and updating software within IoT systems.
Although not all studies directly address improving IoT security using LLM, there is potential for further research and exploration to contribute to IoT security.

\smallskip
\noindent \textbf{\citet{islam2024llmpowered}} proposed an LLM-based tool to patch vulnerable code.
The LLM is trained using semantic reward and reinforcement learning.
It takes C code as input and produces a patched version with fewer or no vulnerabilities.
The study shows successful patching of vulnerabilities that improve the security of IoT devices by preventing initial access points for attackers.
There is potential in this work to ensure that IoT software, possibly at the firmware or operating system level, has minimal vulnerabilities.
The study demonstrates that the patch fixes common and known vulnerabilities, indicating a further potential for LLM to improve in terms of fixes.
The LLM could be trained using open source datasets, such as Automated CVEFixes by ~\citet{bhandari2021:cvefixes}, that focus on IoT.
With datasets specializing in IoT, the LLM can be further trained to patch software that prevents the exploitation of public-facing devices.
LLM could be applied as a tool for automated vulnerability patching to address security weaknesses in the context of the IoT.
The tool is capable of autonomously patching vulnerable code with minimal human input.
Potential for further improvements include additional modules for automated implementation or replacement tasks.
Its impact on security is high, increasing efficiency and effectiveness in software security improvement.

\smallskip
\noindent \textbf{DefectHunter~\citet{wang2023defecthunter}} is another LLM-based implementation for patching vulnerabilities.
It serves a similar purpose to~\citet{islam2024llmpowered} since both use LLMs to repair and patch vulnerable code.
However, DefectHunter differs in its design, utilizing attention models instead of reinforcement learning and semantic rewards.
Both studies demonstrate that current LLMs could effectively patch vulnerable code when given as prompts.
To apply LLM to IoT, it is necessary to incorporate IoT-specific training datasets, such as the QEMU dataset~\citet{zhou2019devign}, Pongo-70B~\citet{Pongo-70B}, and CWE-754 dataset~\citet{NVD}.
This would enable the LLM to understand and patch IoT-specific vulnerabilities and defects.
Potential improvements include modules to optimize processing time and training the model with IoT-specific dataset.
Its impact on security is significant due to the automation of vulnerability patching, which allows faster software review and more efficient code production, leading to a more secure system.
\begin{landscape}
\begin{table}[htbp]
    \centering
    \caption{Analysis on Potentials and Impact of GenAI for IoT Security}
    \label{tab:potentials-genai-iot-sec}
    \catcode`\_=12
    \catcode`\#=12
    \begin{threeparttable}
    \begin{tabular}{|r|l|c|c|c|*{6}{c}|}
    \hline
    \cline{1-11}
    \multirow{2}{*}{#} & \multirow{2}{*}{Mitigation Techniques} & \multirow{2}{*}{Tactics} & \multirow{2}{*}{Techniques Mitigated} & \multirow{2}{*}{LLM Works} 
    & \multicolumn{6}{c|}{Functions and Potentials} \\
    \cline{6-11}
    & & & & & ETD & IAD & RA & RM & DP & IS \\
    \cline{1-11}
    1 & Application Developer Guidance & Persistence    & T0859 & LLMSecGuard~\citet{Kavian2024LLMSG} & \emptycirc & \halfcirc & \fullcirc & \emptycirc & \fullcirc & \halfcirc \\
    2 & Application Developer Guidance & Initial Access & T0819 & LLift~\citet{libugdetection2024} & \emptycirc & \halfcirc & \halfcirc & \fullcirc & \fullcirc & \fullcirc \\
    3 & Application Developer Guidance & Initial Access & T0819 & HuntGPT~\citet{ali2023huntgpt} & \halfcirc & \emptycirc & \fullcirc & \fullcirc & \fullcirc & \fullcirc \\    
    \cline{1-11}
    4 & Exploit Protection & Initial Access & T0866 & LLMind~\citet{cui2024llmind} & \halfcirc & \emptycirc & \fullcirc & \fullcirc & \fullcirc & \halfcirc \\
    5 & Exploit Protection & Privilege Escalation & T0890 & ~\citet{wang2024hybrid} & \halfcirc & \halfcirc & \fullcirc & \emptycirc & \fullcirc & \fullcirc \\
    6 & Exploit Protection & Execution & T0853 & NVISOsecurity~\citet{Raman_2024} & \fullcirc & \emptycirc & \fullcirc & \fullcirc & \fullcirc & \fullcirc \\
    7 & Exploit Protection & Initial Access & T0866 & Cyber Sentinel~\citet{kaheh2023cyber} & \fullcirc & \emptycirc & \fullcirc & \fullcirc & \fullcirc & \fullcirc \\
    \cline{1-11}
    8 & Hardware Installation Limitation & Initial Access & T0847 & VIoTGPT~\citet{zhong2023viotgpt} & \fullcirc & \emptycirc & \fullcirc & \fullcirc & \fullcirc & \emptycirc \\
    \cline{1-11}
    9 & Mechanical Layer Protection & Impact & T0880 & ~\citet{saha2023llm} & \emptycirc & \halfcirc & \halfcirc & \fullcirc & \fullcirc & \fullcirc \\
    \cline{1-11}
    10 & Network Intrusion Prevention & Command And Control & T0869 & BERTIDS~\citet{lira2024} & \halfcirc & \emptycirc & \fullcirc & \emptycirc & \halfcirc & \halfcirc \\
    11 & Network Intrusion Prevention & Command And Control & T0869 & ~\citet{guastalla2023application} & \halfcirc & \emptycirc & \fullcirc & \emptycirc & \halfcirc & \fullcirc \\
    12 & Network Intrusion Prevention & Command And Control & T0869 & SecurityBERT~\citet{ferrag2024revolutionizing} & \halfcirc & \emptycirc & \fullcirc & \emptycirc & \halfcirc & \halfcirc \\
    13 &  Network Intrusion Prevention & External Remote Services & T0886 & IDS-Agent~\citet{li2024idsagent} & \fullcirc & \emptycirc & \fullcirc & \fullcirc & \fullcirc & \fullcirc \\
    \cline{1-11}
    14 & Software Update & Initial Access & T0819 & ~\citet{islam2024llmpowered} & \emptycirc & \fullcirc & \fullcirc & \halfcirc & \fullcirc & \fullcirc \\
    15 & Software Update & Initial Access & T0819 & DefectHunter~\citet{wang2023defecthunter} & \emptycirc & \fullcirc & \fullcirc & \halfcirc & \fullcirc & \fullcirc \\
    \cline{1-11}
    16 & Threat Intelligence Program & Initial Access & T0866 & AttackGen~\citet{Adams_2024} & \fullcirc & \emptycirc & \fullcirc & \fullcirc & \fullcirc &  \fullcirc \\        
    17 & Threat Intelligence Program & Initial Access & T0866 & ~\citet{mcintosh2023harnessing} & \fullcirc & \emptycirc & \fullcirc & \fullcirc & \halfcirc & \fullcirc \\
    \cline{1-11}
    18 & User Training & Initial Access & T0865 & ~\citet{bethany2024large} & \halfcirc & \emptycirc & \fullcirc & \emptycirc & \halfcirc & \halfcirc \\
    19 & User Training & Execution & T0863 & ~\citet{yamin2024} & \halfcirc & \emptycirc & \fullcirc & \halfcirc & \emptycirc & \halfcirc \\
    \cline{1-11}
    20 & Vulnerability Scanning & Initial Access & T0819 & LLM4Vuln~\citet{sun2024llm4vuln} & \emptycirc & \fullcirc & \fullcirc & \fullcirc & \fullcirc & \fullcirc \\
    21 & Vulnerability Scanning & Initial Access & T0819 & AutoAttacker~\citet{xu2024autoattacker} & \fullcirc & \halfcirc & \fullcirc & \fullcirc & \fullcirc & \halfcirc \\
    22 & Vulnerability Scanning & Initial Access & T0819 & ~\citet{tóth2024llms} & \halfcirc & \halfcirc & \fullcirc & \emptycirc & \fullcirc & \halfcirc \\
    23 & Vulnerability Scanning & Initial Access & T0819 & ~\citet{oliinyk2024fuzzing} & \halfcirc & \emptycirc & \fullcirc & \halfcirc & \halfcirc & \halfcirc \\
    24 & Vulnerability Scanning & Initial Access & T0819 & ~\citet{happe2024llms} & \halfcirc & \emptycirc & \halfcirc & \emptycirc & \fullcirc & \emptycirc \\
    25 & Vulnerability Scanning & Initial Access & T0819 & LLMIF~\citet{wangfuzzing2024} & \fullcirc & \emptycirc & \fullcirc & \halfcirc & \halfcirc & \halfcirc \\
    26 & Vulnerability Scanning & Initial Access & T0819 & ChatAFL~\citet{meng2024large} & \fullcirc & \emptycirc & \fullcirc & \fullcirc & \fullcirc & \fullcirc \\
    27 & Vulnerability Scanning & Initial Access & T0819 & FIAL~\citet{androidfuzz2024} & \halfcirc & \emptycirc & \fullcirc & \halfcirc & \fullcirc & \fullcirc \\    
    29 & Vulnerability Scanning & Initial Access & T0819 & ~\citet{fang2024llm} & \fullcirc & \emptycirc & \fullcirc & \halfcirc & \halfcirc & \halfcirc \\
    30 & Vulnerability Scanning & Execution      & T0819 & mGPTFuzz~\citet{Maetal2024} & \emptycirc & \fullcirc & \fullcirc & \emptycirc & \halfcirc &  \halfcirc \\
    31 & Vulnerability Scanning & Execution      & T0819 & PentestGPT~\citet{deng2023pentestgpt} & \fullcirc & \halfcirc & \fullcirc & \fullcirc & \halfcirc & \fullcirc \\
    32 & Vulnerability Scanning & Initial Access & T0819 & Net-GPT~\citet{10419242} & \fullcirc & \emptycirc & \fullcirc & \halfcirc & \fullcirc & \halfcirc \\
    33 & Vulnerability Scanning & Initial Access & T0819 & ~\citet{Happe_2023} & \fullcirc & \emptycirc & \fullcirc & \fullcirc & \fullcirc & \halfcirc \\
    34 & Vulnerability Scanning & Initial Access & T0819 & ~\citet{yang2023iot} & \emptycirc & \halfcirc & \fullcirc & \emptycirc & \fullcirc & \halfcirc \\
    \hline
    \end{tabular}
    \begin{tablenotes}[para]
        ETD: External Threat Detection.
        IAD: Internal Anomaly Detection.
        RA: Response Automation.
        RM: Research Maturity.
        DP: Development Potential.
        IS: Implact on Security.        
    \end{tablenotes}
    \end{threeparttable}
\end{table}
\end{landscape}
\smallskip
\noindent \textbf{LLift~\citet{libugdetection2024}} is another LLM-based tool for patching vulnerabilities.
As mentioned in Section~\ref{subsec:adg}, it is used to identify UBI bugs and could be applied during the development phase or to identify vulnerabilities in a running Linux kernel.
LLift performs a static analysis to find unpatched vulnerabilities, allowing users to patch them.
Despite a precision of 50\%, LLift is effective in identifying vulnerabilities, making it a useful tool for patching them.
This LLM implementation could help mitigate IoT attacks by updating software to patch vulnerabilities.
\subsection{Threat Intelligence Program}
This section outlines the research conducted on LLMs to mitigate attacks by developing threat intelligence policies.
These mitigation efforts encompass various approaches, including the formulation of security policies for organizations and incident response plans.

\smallskip
\noindent \textbf{AttackGen~\citet{Adams_2024}} is an LLM-powered incident response tool that helps organizations prepare for cyber attacks by understanding possible attack vectors.
It automatically generates these scenarios based on industry type, attack vectors, and organization size.
AttackGen uses these parameters to create detailed incident response scenarios, with OpenAI as the default model.
In its default setting, AttackGen could generate general incident response plans and evaluation metrics.
It is a viable tool for generating threat intelligence to mitigate attacks.
Extending it to IoT systems would involve modifying the prompt to focus on the IoT context.
This could help generate specific incident response plans for IoT.
AttackGen helps prevent external threats by providing incident reports and playbooks for user training, addressing potential attacks and protection methods with potential for further specialization and contextual relevance. 
Its impact on security is high due to its pioneering role in automated report generation and playbook creation, significantly affecting the field of security.
Section~\ref{sec6} discusses a case study on AttackGen with the necessary modifications for IoT implementation.

\smallskip
\noindent \textbf{\citet{mcintosh2023harnessing}} investigated whether GPTs could generate better cyber security policies than humans.
Using a ransomware attack as a case study, they found that GPTs outperformed humans in terms of completeness, effectiveness, and efficiency.
GPTs scored higher on these metrics, indicating that they could generate more secure policies.
Although not yet tested in IoT, the results suggest that GPTs could also be effective in this context.
Transfer learning could also further enhance the LLM focus on IoT, potentially leading to GPTs outperforming human policies in this area.
Its impact on security is significant due to its novelty and effectiveness.
\subsection{User Training}
This section explains how LLMs have been used in studies to improve human skills in IoT security, similar to concepts in generic cyber security.
LLM applications focus on improving awareness of common exploitation methods, such as phishing emails and social engineering.

\smallskip
\noindent \textbf{\citet{bethany2024large}} implemented LLM to generate spear-phishing emails to gain access to the system.
Over 11 months, they found that more than 10\% of the staff in an educational organization were vulnerable to LLM-generated attacks and gave out their credentials.
The study concluded that user training and awareness are needed to prevent such attacks.
The study also resulted in an application to defend against LLM-generated phishing emails, achieving an F1 score of 98.96\%.
In IoT systems, user training is crucial to prevent spear-phishing.
Training and fine-tuning the LLM could create a program based on its e-mail detection capabilities.
Common signs of LLM-generated emails could be compiled into a database to help users identify and avoid such attacks.
This demonstrates the potential of LLM in detecting and training users to mitigate spear-phishing attacks.
The proposed tool mimics attacker methods and functions as a protective tool to prevent external threats.
The LLM operates autonomously with minimal human input, focusing solely on email generation.
Its main goal is to improve the efficiency of phishing content generation.
Although it contributes to security training, its impact is limited due to the well-established nature of phishing awareness and existing preventive measures.

\smallskip
\noindent \textbf{\citet{yamin2024}} emphasized the need for personnel training to gain experience during cyber attacks.
Since real attacks are hard to predict, cyber exercises are used for training.
The authors created a scenario generation tool using LLM to produce exercise scenarios based on various criteria, following the concept of digital game-based learning~\citet{digitalgamebasedlearning}.
This tool could generate scenarios for both known and emerging security issues.
In IoT systems, specific prompts can be used to generate threats as exercise scenarios.
This demonstrates the potential of LLM as a tool for IoT security training, enhancing the capabilities of IoT security personnel.
For example, a compromised smart home scenario could be simulated as an exercise.
The tool does not generate playbooks or handle OS/software anomalies, focusing solely on human training.
The LLM autonomously creates scenarios based on user queries.
Although it improves training efficiency, its impact on security is limited, as it does not provide implementation guidance.
The tool addresses a developing field in security, with potential improvements mainly in content generation effectiveness.

\subsection{ Vulnerability Scanning}
This section explains how LLMs could enhance IoT security by identifying and fixing vulnerabilities in devices.
It covers both finding and fixing these vulnerabilities, including security testing during development to prevent issues before deployment.

\smallskip
\noindent \textbf{LLM4Vuln~\citet{sun2024llm4vuln}} studied the reasoning abilities of LLM to identify vulnerabilities and understand the key components affecting this process.
The authors focused on smart contracts and identified knowledge retrieval, tool invocation, prompt schemes, and instruction following as critical factors.
The experimental results showed that knowledge retrieval is crucial and that GPT-4 performed best among LLMs.
Using LLM4Vuln, nine zero-day vulnerabilities in bug bounty programs were identified.
This work suggests that applying LLM4Vuln to IoT could help improve LLM’s ability to identify vulnerabilities in IoT security.
LLM4Vuln operates with minimal user interaction, autonomously discovering vulnerabilities based on its training.
It addresses the evolving field of vulnerability exploration, indicating the potential for further development.
Future improvements could include an automated task executor to fix vulnerabilities and additional contextualization for specific security fields.
This autonomous agent could improve system security, particularly in the IoT context.

\smallskip
\noindent \textbf{AutoAttacker~\citet{xu2024autoattacker}} uses LLM to automate attack launching, serving as a red-team tool.
It is a jailbroken LLM that could execute complex tasks such as lateral movement and obtaining credentials on Windows and Linux platforms, leveraging GPT-4 and Metasploit.
It successfully executed all benchmark attack tasks.
Fine-tuning AutoAttacker with a dataset like CICIoT2023, which includes various attacks targeting IoT devices, could create a specialized LLM.
This would enable AutoAttacker to focus on IoT-specific attack scripts and enhance security testing and defense mechanisms for IoT.
AutoAttacker could successfully exploit known vulnerabilities, highlighting areas for defense improvement.
It addresses the evolving field of defense capabilities, with potential for further improvement to handle more complex tasks and discover new vulnerabilities.
AutoAttacker's impact on the security field is significant, though limited to discovering known vulnerabilities.
It improves defense through vulnerability discovery, but lacks the ability to find new ones.

\smallskip
\noindent \textbf{\citet{tóth2024llms}} implemented an LLM to scan and find vulnerabilities in web environments, focusing on PHP code and common web attacks such as XSS and SQL injection.
The authors used GPT-4 to generate PHP code and GPT-3.5 with static code analysis to find vulnerabilities.
The LLM identified 78\% of static file upload vulnerabilities, 50.15\% of prepared statement vulnerabilities, 38\% of code audit vulnerabilities, 11.16\% of XSS or SQL injection vulnerabilities, and 8\% of vulnerable code manually.
This approach could be adapted for IoT by modifying the code generation and classification steps.
Although IoT software differs from web applications, this framework shows the potential to find vulnerabilities in AI-generated code, helping mitigate attacks.
The tool operates with minimal human input, focusing on scripts and code related to web applications.
Broadening its scope to include different contexts could improve its capabilities.
This could involve adding other datasets and developing an automated executor based on identified vulnerabilities.
The tool's impact on security is limited due to its focus only on web-based security.

\smallskip
\noindent \textbf{\citet{oliinyk2024fuzzing}} present a new method for creating a security testing tool, specifically a fuzzer.
Their study used a trained LLM to fuzz BusyBox, a tools suite that combines many Unix utilities.
The LLM effectively crashed the environment and identified weaknesses without traditional fuzzing methods.
This shows that LLM could test the security of Unix-based systems, which is relevant for IoT devices, since they often use Unix-like operating systems (e.g. embedded Linux).
The study suggests that LLM could automate and streamline the fuzzing process for IoT devices, making security testing more efficient.
This allows more time for additional security tests before deployment.
Therefore, this approach could improve the effectiveness and efficiency of security testing for IoT devices.
The tool generates automated inputs for fuzzing with minimal human input, continuously testing until new vulnerabilities are found.
Although the field is mature, further research could improve the tool by expanding its application beyond embedded Linux systems to other types of systems.
The tool's impact on security is limited to fuzzing, increasing the efficiency of vulnerability discovery without providing solutions to patch them or protecting the system afterwards.

\smallskip
\noindent \textbf{\citet{happe2024llms}} use LLM for privilege escalation, functioning as a red-team tool in a controlled environment.
They focus on escalating privilege once inside the system.
Fine-tuned LLMs, specifically Llama-2 and GPT-4, were tested on a Linux privilege-escalation benchmark, with GPT-4 performing better.
The LLM runs commands to escalate the attacker’s privilege.
Limitations include the LLM running the same commands repeatedly.
In IoT testing, this LLM implementation could help create defense mechanisms to identify and block LLM-generated commands attempting privilege escalation.
The tool fully automates the privilege escalation process with initial queries and inputs, increasing efficiency and effectiveness.
However, it repeats the same query without human intervention, which is a weakness.
Privilege escalation is a well-researched field, and while the tool improves process efficiency, its impact on security is limited to this specific area.
The development potential includes overcoming the repetition issue to enhance automation.

\smallskip
\noindent \textbf{\citet{wangfuzzing2024}} improved the efficiency and effectiveness of software testing using LLM for fuzzing.
The authors addressed limitations such as unknown message formats, unresolved dependencies, and lack of testing evaluations.
The proposed model, LLMIF, uncovered 11 vulnerabilities, including eight new ones, in Zigbee devices.
This makes LLMIF useful for discovering vulnerabilities in IoT systems, helping security experts identify and patch them in the future.
LLMIF autonomously generates input to test the defenses of the IoT system with minimal user intervention, achieving relatively complete automation of the fuzzing process.
The tool addresses the well-established field of fuzzing, with potential for further development through LLM.
LLMIF's impact on the security field is significant within the fuzzing niche, as it can discover new vulnerabilities more effectively than humans.
Despite its niche focus, its ability to find new vulnerabilities highlights its impact.

\smallskip
\noindent \textbf{ChatAFL~\citet{meng2024large}} is an LLM-based protocol fuzzer to test protocol implementation correctness and vulnerabilities.
A protocol fuzzer is defined as a tool that generates message sequences following the required structure and order of a protocol.
ChatAFL performed faster and covered more branches than other benchmark fuzzers such as AFLNet~\citet{9159093} and NSFuzz~\citet{NSFuzz2023}.
The addition of LLM to the protocol fuzzing improved efficiency and coverage.
Although not tested on IoT, ChatAFL has potential as a security testing tool if fine-tuned with an IoT dataset.
This could enable ChatAFL to perform protocol fuzzing on emerging IoT protocols such as Matter, serving as a red-team tool for vulnerability scanning.
ChatAFL autonomously generates messages in the given protocol format, requiring minimal user intervention.
It addresses the emerging field of protocol fuzzing, which has gained attention since 2019.
The development potential for ChatAFL is tied to the evolving field of protocol fuzzing.
Further training and improvement depend on advancements in the field due to the lack of a standardized dataset.
ChatAFL's impact on the security field is significant, particularly in IoT, as it enhances the efficiency and effectiveness of protocol fuzzing through automation.

\smallskip
\noindent \textbf{FIAL~\citet{androidfuzz2024}} is an implementation of LLM as a fuzzing tool for IoT devices.
It uniquely employs an Android device for execution, combined with taint analysis results to generate suitable trigger functions for fuzzing.
The Android device extracts network packets, sends them to the LLM and data analyzer, and receives a crash code to test on the IoT system.
If the system crashes, a successful vulnerability exploitation is identified.
The experiment results identified 14 vulnerabilities, including 3 injection and 11 overflow vulnerabilities, of which 5 being new.
This demonstrates a unique LLM implementation using an Android device for fuzzing and finding vulnerabilities in IoT systems, successfully implementing a tool for vulnerability scanning.
FIAL is able to discover new vulnerabilities autonomously, but is limited to network attack vectors.
It requires minimal human intervention, with the main interface being an Android device, which limits the types of input and commands.
Fuzzing is a mature research field but continues to evolve with software advancements.
In IoT security, automation, and research through LLM, as seen with FIAL and CHATAFL, improve the fuzzing process.
The development potential for FIAL includes creating a more mobile and stealthy device for running tests.
There is also potential for developing a similar tool for iOS devices.
FIAL's impact on security is significant, especially in fuzzing and testing.
The use of an Android device allows for stealth testing and fuzzing, which could lead to new research directions for prevention and detection.

\smallskip
\noindent \textbf{\citet{fang2024llm}} demonstrated a method for an automated attacker using LLM to secure IoT systems.
The study showed that an LLM trained with GPT-4 executed 87\% of known attacks when given the CVE description, but only 7\% without it.
This highlights the potential of LLM tools for security testing.
With further research, this tool could automate vulnerability exploitation in cyber exercises or training.
Additionally, there is potential for LLM to eliminate one-day vulnerabilities by patching them immediately upon discovery.
It effectively addresses external threats and could find new vulnerabilities without CVEs.
The tool autonomously exploits vulnerabilities with given CVEs and to a limited degree without them, fulfilling its purpose with minimal human intervention.
Simulated attack tools are a mature research topic, and this tool improves automation and vulnerability exposure capabilities.
The development potential includes improving the discovery of new vulnerabilities without CVEs.
The impact of the tool on security is limited to its automated attack capabilities.

\smallskip
\noindent \textbf{mGPTFuzz~\cite{Maetal2024}} is a first-of-its-kind Matter fuzzer to find bugs and vulnerabilities in Matter-compatible IoT devices.
The authors leverage LLM to transform the human-readable specification, over a thousand pages, to machine-readable information in the form of finite state machines (FSMs).
It is a blackbox fuzzing tool and mGPTFuzz is able to find stateful, non-crash and crash bugs.
mGPTFuzz was evaluated with 23 Matter devices leading to the discovery of 147 new bugs, including 61 zero-day vulnerabilities and three CVEs.
While the fuzzer itself is limited to Matter-compatible devices, this work has significant impact on IoT security due to increasing support and adoption of Matter-certified smart home devices.

\smallskip
\noindent \textbf{PentestGPT~\citet{deng2023pentestgpt}} used LLM, such as GPT-4 and Bard, to automate the penetration testing process.
It leverages the knowledge of pre-trained LLMs to conduct these tests.
PentestGPT was tested against benchmark attacks and divided tasks to compare LLM performance.
It was 228.6\% more effective than other LLMs and applicable to real life challenges.
Fine-tuned PentestGPT significantly improved task execution and results.
PentestGPT also enhances the penetration testing process by relating steps for a more effective execution.
This study demonstrates a practical and effective implementation of LLM for automated penetration testing.
The tool automates the penetration testing process with minimal human input, outperforming human users in benchmark tests.
Penetration testing is a constantly evolving field that offers potential for improvement in defense mechanisms.
The development potential includes further training for different contexts such as IoT or ICS.
Its impact on the security field is significant as it improves the effectiveness of protection mechanisms through efficient vulnerability discovery.

\smallskip
\noindent \textbf{Net-GPT~\citet{10419242}} used LLM as a red-team tool to launch automated man-in-the-middle attacks on unmanned aerial vehicles (UAVs).
It claims an efficacy of more than 90\% in hijacking UAVs after fine-tuning Llama GPTs and more than 70\% with other smaller LLMs.
Net-GPT assumes that the attacker has access to the network, compromises a UAV, and intercepts communications between the UAV and ground control.
For IoT, a specific man-in-the-middle attack dataset is needed to fine-tune the LLM, allowing it to learn IoT-specific behaviors.
In addition, a benign IoT device must be compromised to act as a foothold to observe, modify, and compromise the IoT network, similar to the UAV implementation.
Net-GPT mimics network packets and act as a man-in-the-middle between two devices in the system.
The tool addresses the established field of man-in-the-middle attacks, which already has early detections and mitigations.
Development potential includes further contextualization and expanding the scope of the LLM's application.
Net-GPT's impact on security is specific to man-in-the-middle attack vectors, exposing vulnerabilities to this type of attack but not others.

\smallskip
\noindent \textbf{\citet{Happe_2023}} addressed both high-level penetration testing planning and low-level execution using LLM.
The LLM gained root privilege and obtained passwords on a compromised Linux machine using the ``sudo -l'' command and reading ``/etc/passwd''.
It could also create a reverse shell, though less consistently.
This suggests that LLM can automate penetration testing tasks and planning.
Given the experimental environment was a compromised Linux system, commonly used by IoT devices, it has potential for automated penetration testing of IoT systems running Linux.
The tool operates with minimal human response, increasing the efficiency of the penetration testing process.
Penetration testing is a mature field that continues to grow with advances in security.
The development potential includes improving the model to increase the variety and consistency of simulated attacks for vulnerability exposure.
Another direction is to contextualize attacks for the IoT field, extending current tests on Linux systems.
The tool's capability is limited to network-based penetration testing, restricting the attack vectors that could be tested.

\smallskip
\noindent \textbf{\citet{yang2023iot}} studied the use of LLM and static code analysis to identify and constrain vulnerabilities in IoT systems through user prompts.
Their study showed a 66.67\% success rate in identifying vulnerability types with an average of 9 prompts and an 83.33\% success rate in identifying specific code lines with an average of 4 prompts.
Prompt engineering was found to be at least 60\% effective in both tasks.
This work highlights the potential of prompt engineering to scan for vulnerabilities in IoT systems.
Further research could improve the effectiveness of this approach in identifying vulnerabilities.
The tool autonomously executes tasks to constrain the type of vulnerability with minimal user intervention.
It functions within the mature research field of static code analysis, automating the process to increase security.
Potential improvements include automated execution of vulnerability fixing and enhanced detection capabilities for more efficient and effective queries.
The impact of the tool on security is limited to increasing the effectiveness and efficiency of static code analysis, as it does not handle other protection or detection methods, limiting its overall impact.
\section{Case Study}
\label{sec6}
This section discusses three case studies: AttackGen~\citet{Adams_2024}, NVISOsecurity~\citet{Raman_2024}, and ChatIoT~\citet{dong2024chatiot} along with their potential implementation for IoT security.
AttackGen, an LLM-based Incident Response Plan Generator, shows potential in generating IoT-related incident response plans.
NVISOsecurity, using AutoGen~\citet{wu2023autogen} as its backend and Caldera agents as executors, demonstrates potential through its performance in executing complex tasks with pre-defined prompts.
ChatIoT is an LLM-based assistant designed to facilitate IoT security and threat intelligence by leveraging the versatile property of retrieval-augmented generation (RAG) illustrating a promising new direction in integrating the advanced language understanding and reasoning capabilities of LLM with fast-evolving IoT security information.
We provide details on the case studies and their potential implementation for IoT security in the following sections.
\subsection{AttackGen}
%
%
A case study on AttackGen~\citet{Adams_2024} was conducted to evaluate its potential to implement LLM in the mitigation of threat intelligence programs.
The inputs to AttackGen are the LLM model, industry, and the size of a hypothetical organization.
To assess AttackGen’s potential in creating incident response plans for IoT systems, the following modifications were made.
Firstly, the MITRE ATT\&CK Enterprise matrix was replaced with the ICS matrix to support the generation of incident response scenarios tailored to industrial and critical infrastructure environments, as the closest matrix for IoT.
The prompt template was then modified to focus on incident response for IoT.
Subsequently, the assistant module was specified to refine and identify critical IoT devices within the organization. 
After the modifications, AttackGen was used to generate an incident response plan focused on IoT with the GPT-4o model, targeting the energy sector with the Dragonfly group~\citet{Dragonfly}.

In this case study, human experts evaluated the relevance, clarity, and specificity of the generated plan.
Relevance was measured by how much the plan related to the Mitigation Technique for the specific threat group and business sector.
Clarity was measured by the plan's ability to create a test plan that human testers could follow, with detailed steps for testing Mitigation Techniques.
Specificity was measured by the plan's ability to specify vulnerable devices within the system, including device type, model, brand, or system architecture.
An extra prompt: ``Can you specify this in the context of a possible IoT-powered PLC that is critical to the company and is connected to the Internet?'' was added to the AttackGen Assistant module in order to generate a more refined plan.

AttackGen generated coherent plans from both the original and edited prompts without additional domain-specific IoT training.
Both plans focused on ICS Mitigation Techniques for the threat group, without mentioning specific attack vectors, for example, a vulnerable HVAC system or compromised machine.
The generated response provides an overview of potential attacks by the Dragonfly group through a theoretical supply chain compromise, indicating known attack methods.
Although relevant for incident response testing, both plans lacked focus on the specified energy domain.
Neither plan provided clear and specific instructions to test possible vulnerabilities. The response mentioned steps, triggers, and responses testing without elaborating on domain-specific techniques.
Both plans received scores of 3 out of 5, as neither clearly showed specific steps and instructions for humans to follow.

The refined response generated a plan more related to the IoT context, although it did not mention specific devices or system architecture.
The original plan as provided in Figure~\ref{fig:attackgen-default} focused on ICS Mitigation Techniques related to Dragonfly without mentioning IoT or specific devices.
The refined plan in Figure~\ref{fig:attackgen-iot}, using the ICS Mitigation Techniques as a base, included IoT.
This shows that the added prompt helped to make the plan more specific.
Without domain-specific training, the LLM model could only create generic incident response plans as a starting point for further refinements.
Even with details of the possible attacker and added focus on IoT, the pre-trained GPT-4o could not generate a tailored incident plan for an IoT context.
This shows the potential for improvement and research to enhance the specificity of the generated content, demonstrating the potential for LLMs to be used as tools in the creation of threat intelligence programs for attack mitigation.
\subsection{NVISOsecurity Cyber Security Agent}
%
%
In this case study, we focus on NVISOsecurity's ability to execute commands to list the privileges available for the Caldera agent.
The tool operates by specifying a series of prompts (actions) to the \texttt{task\_coordinator} agent, which then transmits them to the \texttt{caldera\_agent}.
An environment for Caldera was first created with two default agents deployed to simulate communication through TCP (Manx) and HTTPS contacts (Sandcat).
These agents were kept alive while a pair of LLMs were started.

Using the predefined commands, \texttt{HELLO\_CALDERA}, \texttt{DETECT\_} \texttt{AGENT\_PRIVILEGES}, and \texttt{TTP\_REPORT\_TO\_TECHNIQUES} were executed to gather necessary information.
In \texttt{HELLO\_CALDERA}, the LLM pair could access PowerShell or the terminal and run a command to display a text box with a string.
This functionality serves as a simple prototype for further commands that the worker agent could execute.
\texttt{COLLECT\_CALDERA\_INFO} tasked the LLM pair with collecting user privilege information for the Caldera process, which runs at an administrative level.
This allowed the worker agent to execute a command to view the user privileges of the Caldera agent process.
\texttt{TTP\_REPORT\_TO\_TECHNIQUES} download a file from Microsoft, change its format, and identify MITRE techniques.
The worker agent's capabilities include running complex commands to access storage, identify, and format MITRE techniques within documents.

NVISOsecurity demonstrates the potential of LLMs to protect IoT systems through exploit protection by accessing administrative-level commands and obtaining privileged information.
In the IoT context, this means that connected devices could be continuously monitored and protected from anomalous processes.
By adding a custom command to NVISOsecurity and setting up the Caldera OT plugin as the agent, the LLM could block the execution of commands at the administrative level.
This suggests that NVISOsecurity could serve as a semi-autonomous agent, continuously observing and protecting the system from malicious processes.
Additionally, the \texttt{TTP\_REPORT\_TO\_TECHNIQUES} command shows that LLMs can identify specific items, such as MITRE techniques, from text documents.
This indicates the potential for LLMs to act as report generators, taking logs from the worker agent to prevent anomalous processes in an IoT context.

\subsection{ChatIoT}
This case study focuses on the capability of ChatIoT to provide reliable, relevant and technical answers to different types of users about IoT security~\citet{dong2024chatiot}.
This IoT security and threat assistant is built upon RAG, which retrieves external IoT security and threat data and feeds them into LLM to improve the quality of answers.
At a high level, ChatIoT integrates multiple datasets from different sources, including IoT vulnerabilities and exploits, MITRE ATT\&CK TTPs, threat reports, and cyber security labels of IoT devices.
These datasets are pre-processed into the system, and during the service, i.e., when a user submits a query, only the relevant data are retrieved into LLM to synthesize the final results. 
Moreover, Dong et al. proposed a data processing toolkit to convert datasets of various formats into documents for LLM processing and user role-specific prompts to dynamically retrieve data and generate answers aligned with users' expertise levels.

We rely on LLM-as-judges to evaluate the reliability, relevance, technicality, and user-friendliness of the answers.
Reliability measures the trustworthiness of each answer, relevance assesses how well the answer addresses the specific question and meets the user’s needs, technicality is used to judge the appropriateness and precision of technical language, including IoT research, standards, protocols, and relevant technical aspects, and user-friendliness determines how easy the answer is to comprehend, focusing on clarity for the user’s role and background.
An extra prompt: "\textbf{\textit{Instructions}}: \textit{1) Criteria: The descriptions about Reliability, Relevance, Technical, and User-friendliness.
2) Score: \romannumeral1) Provide a score for each answer across the metrics above. Scores should range from 0 to 5, with 5 being the highest and 0 being the lowest; \romannumeral2) Scores should reflect how well each answer meets the criteria, particularly in alignment with the user role's background and needs.
3) Output Format: Present a table that includes the names of all answers and their scores for each metric. You can score differently for different metrics.}"
was added to guide the LLM-based judge for evaluation.

IoT security is constantly evolving with new vulnerabilities, exploits, and security protocols.
ChatIoT automatically extracts the latest information from different sources to aid in LLM's understanding and reasoning.
At the same time, retraining or fine-tuning LLM is costly, resource intensive, and LLMs become out-of-date quickly.
ChatIoT provides a versatile and effective approach to combining LLM's capabilities with up-to-date IoT security and threat intelligence. 
In conclusion, ChatIoT shows the potential of using large language models to effectively facilitate IoT security assistance to various key users of IoT ecosystems in an understandable and actionable manner to provide better IoT security guarantees.
\section{Future Directions and Open Research Problems}
\label{sec:future_works}
This section discusses open research directions and practical applications of GenAI to enhance IoT security.
To categorize potential research directions, we use ICS Mitigation Techniques as a starting point, identifying techniques not addressed in existing sources.
Subsequently, we explore methods to address these topics.

\smallskip
\noindent \textbf{Access Management (M0801): } A potential research direction to address this issue is training an LLM to autonomously enforce authorization policies and decisions, ensuring user identification and verification.
Current LLM implementations have not directly addressed authorization policies for IoT systems, making it an open research problem.
The current capabilities of IoT devices may not support this function, requiring external support within or connected to the network to enforce user authorization and prevent compromise.
One possible approach is to use communications as inputs to LLM and verify the source and destination of IoT devices within the ecosystem.

\smallskip
\noindent \textbf{Encrypt Network Traffic (M0808): } To establish secure communication between PLCs, LLMs could execute complex tasks to improve communication security.
This research leverages the ability of LLMs to perform complex tasks based on prompt requirements, as demonstrated by NVISOsecurity~\citet{Raman_2024}.
LLMs could execute various lightweight cryptography algorithms, adding controlled randomness to network encryption.
For instance, the LLM might execute RSA encryption for one message and elliptic curve encryption for the next.
This dynamic approach ensures secure communication, as the encryption algorithm changes based on the LLM's directives.

\smallskip
\noindent \textbf{Operating System Configuration (M0928): } While existing LLMs could configure and secure system configurations, there is no implementation that automates configuration across all devices within an IoT ecosystem.
A practical application could involve using LLM as an agent to execute commands and change configurations based on available devices.
The LLM, connected to the network, would identify devices and send device-specific commands to modify configurations across different operating systems, improving security.
This approach presents a potential research avenue for using LLMs to improve the security of the IoT system through configuration changes.

\smallskip
\noindent \textbf{Supply Chain Management (M0817): } A potential research direction involves the use of LLM to manage the supply chain of IoT devices.
Management here means coordinating according to a given policy to minimize the risks of supply chain compromises.
Studies show that LLMs could generate policies and cyber-exercise scenarios, indicating their ability to produce a detailed, domain-specific response to enhance security. 
In supply chain management, LLMs could generate policies to verify the validity of the supplier. 
Trained with vast datasets, LLMs could leverage historical data on entities, including potential IoT device suppliers.
This capability allows LLMs to create purchase or usage policies and recommendations to improve the security of the IoT ecosystem.
A study by~\citet{li2023large} optimized the management process but not the security of the supply chain.
Using this study as a starting point, there is potential to fine-tune the model to include security considerations, such as vendor risk profiles, hardware design, and supply chain attacks, allowing LLMs to mitigate risks effectively.

\smallskip
\noindent \textbf{Validate Program Inputs (M0818) : } LLMs could validate user input on IoT devices to determine their validity.
They take advantage of vast knowledge to identify patterns in user inputs.
By learning patterns of valid and invalid behaviors, LLMs could potentially replicate actions to validate inputs and learn valid user behaviors.
If input comes from invalid IP addresses, they could be classified as malicious.
However, if the input comes from a valid device but does not match the learned behaviors, a trained LLM could identify and reason whether the input is valid.
~\citet{munley2024llm4vv} developed a compiler validation test suite using LLMs to generate test cases.
This concept could be adapted to create behavioral test cases for LLMs to validate program input. 
Research could explore using LLMs to differentiate valid inputs based on user behavior.
\section{Key Takeaways} \label{sec:conclusions}
In conclusion, current GenAI implementations address IoT security improvements.
LLMs have been used in various ways, such as finding vulnerabilities and creating cyber exercise scenarios, to improve cyber security and directly improve IoT security.
Although these implementations are robust, there are inherent limitations, such as the lack of a publicly available repository for LLM security issues.
However, this does not prevent LLMs from being further research to address identified gaps, potentially securing the IoT ecosystem, its network, and its human aspect.
The 33 state-of-the-art works serve as starting points for further research. 
Accompanied with three case studies, this survey paper inspires the application of GenAI to secure IoT systems, be it hardware, software, or network security.
\section*{Acknowledgements}
This research is supported by the National Research Foundation, Singapore, under its National Satellite of
Excellence Programme “Design Science and Technology for Secure Critical Infrastructure: Phase II” (Award No:
NRF-NCR25-NSOE05-0001). Any opinions, findings and conclusions or recommendations expressed in this material
are those of the author(s) and do not reflect the views of National Research Foundation, Singapore.
%
%
%
%
%


\bibliographystyle{elsarticle-num-names} 
\bibliography{references}
\appendix
\section{Evaluation of State-of-the-Art Works}
We discussed each state-of-the-art work according to the six capabilities as described in Section~\ref{sec:detection}.
Table~\ref{tab:potentials-details} provides justifications and evaluation results for 33 state-of-the-art works.
\begin{landscape}
\begin{table}
    \centering
    \caption{Evaluation and Justification of State-of-the-Art Works}
    \label{tab:potentials-details}
    \catcode`\_=12
    \catcode`\#=12
    \begin{threeparttable}
    \begin{tabular}{|r|c|c|c|l|}
    \hline    
    \thead[c]{#} & \thead[c]{GenAI / LLM Works} & Capability & Evaluation & \thead[c]{Justification} \\
    \cline{1-5}
    \multirow{6}{*}{1} &
    \multirow{6}{*}{\makecell[c]{LLMSecGuard\\\citet{Kavian2024LLMSG}}}
      & ETD & \emptycirc & Focuses on secure code generation but does not detect external threats. \\
    & & IAD & \halfcirc & Detects internal vulnerabilities but does not identify backdoors, network anomalies, etc. \\
    & & RA  & \fullcirc & Automates analysis and patching process with minimal user input. \\
    & & RM  & \emptycirc & Secure coding is a well-established field with limited opportunities for new research. \\
    & & DP  & \fullcirc & Potential for further development through fine-tuning in specific fields. \\
    & & IS  & \halfcirc & Limited to addressing internal code vulnerabilities. \\
    \cline{1-5}
    \multirow{6}{*}{2} &
    \multirow{6}{*}{\makecell[c]{LLift\\\citet{libugdetection2024}}}
      & ETD & \emptycirc & Detects UBI bugs but does not identify external threats. \\
    & & IAD & \halfcirc & Does not address the detection of broader internal anomalies. \\
    & & RA  & \halfcirc & Does not incorporate patching, limiting automation. \\
    & & RM  & \fullcirc & Research in UBI bug detection is still in progress, leaving room for further study. \\
    & & DP  & \fullcirc & Potential for automated patching and real-time detection. \\
    & & IS  & \fullcirc & Impactful as UBI bugs could lead to privilege escalation and information leakage. \\
    \cline{1-5}
    \multirow{6}{*}{3} &
    \multirow{6}{*}{\makecell[c]{HuntGPT\\\citet{ali2023huntgpt}}}
      & ETD &  \halfcirc & Detects external threats but only analyzes reports, not real-time threats. \\
    & & IAD & \emptycirc & Does not support internal anomaly detection. \\
    & & RA  & \fullcirc & Generates detailed reports and explanations with an interactive dashboard. \\
    & & RM  & \fullcirc & More IoT datasets for fine-tuning with many new works in this field. \\
    & & DP  & \fullcirc & Extensible towards live intrusion detection and internal anomaly detection. \\
    & & IS  & \fullcirc & Automates forensic analysis and reduces the time required to classify external threats. \\
    \cline{1-5}
    \multirow{6}{*}{4} &
    \multirow{6}{*}{\makecell[c]{LLMind\\\citet{cui2024llmind}}}
      & ETD & \halfcirc & Designed for automated task execution with limited external threat detection capabilities. \\
    & & IAD & \emptycirc & Does not support internal anomaly detection. \\
    & & RA  & \fullcirc & Fully automated in executing tasks, requiring no user input. \\
    & & RM  & \fullcirc & Research in task automation is still evolving. \\
    & & DP  & \fullcirc & Development potential to expand into security-specific tasks such as automated IP blocking. \\
    & & IS  & \halfcirc & Limited to simple network and physical security tasks. \\
    \cline{1-5}
    \multirow{6}{*}{5} &
    \multirow{6}{*}{~\citet{wang2024hybrid}}
      & ETD & \halfcirc & Prevents external threats by identifying UPR variables but no active detection or mitigatation. \\
    & & IAD & \halfcirc & Provides limited internal anomaly detection but does not address broader vulnerabilities. \\
    & & RA  & \fullcirc & Capable to identify UPR variables and prompt users with minimal human intervention. \\
    & & RM  & \emptycirc & Little room for further improvements. \\
    & & DP  & \fullcirc & Development potential includes specialization for IoT security and automated patching. \\
    & & IS  & \fullcirc & Impactful since it prevents privilege escalation through UPR variable identification. \\
    \hline
    \end{tabular}
    \end{threeparttable}
\end{table}
\end{landscape}
\begin{landscape}
\begin{table}
    \centering
    \catcode`\_=12
    \catcode`\#=12
    \begin{threeparttable}
    \begin{tabular}{|r|c|c|c|l|}
    \hline    
    \multirow{6}{*}{6} &
    \multirow{6}{*}{\makecell[c]{NVISOsecurity\\\citet{Raman_2024}}}
      & ETD & \fullcirc & Detects external threats by simulating various attacks based on the MITRE framework. \\
    & & IAD & \emptycirc  & Focuses entirely on external attack simulation and response. \\
    & & RA  & \fullcirc & Fully automates security testing by executing attack-defense scenarios with minimal human input. \\
    & & RM  & \fullcirc & The field is continuously evolving with MITRE and Metasploit. \\
    & & DP  & \fullcirc & Development potential towards real-time threat response and IoT security. \\    
    & & IS  & \fullcirc & Major impact on cybersecurity by automating attack-defense simulations and improving security standards. \\
    \cline{1-5}
    \multirow{6}{*}{7} &
    \multirow{6}{*}{\makecell[c]{Cyber Sentinel\\\citet{kaheh2023cyber}}}
      & ETD & \fullcirc & Detects external threats by retrieving and blocking IP addresses to prevent exploitation. \\
    & & IAD & \emptycirc & No evidence that it detects internal anomalies, though it may have potential for such functionality. \\
    & & RA  &  \fullcirc & Fully automated, executing security tasks with minimal user input. \\
    & & RM  &  \fullcirc & Automated security task execution with LLMs is still an active research field. \\
    & & DP  &  \fullcirc & Development potential for further IoT security applications and penetration testing. \\
    & & IS  &  \fullcirc & Simplifies security tasks for non-experts and improves automation. \\
    \cline{1-5}
    \multirow{6}{*}{8} &
    \multirow{6}{*}{\makecell[c]{VIoTGPT\\\citet{zhong2023viotgpt}}}
      & ETD & \fullcirc & Detect external threats through visual analysis, identifying physical attacks and anomalous behavior. \\
    & & IAD & \emptycirc & Solely focuses on physical security. \\
    & & RA  & \fullcirc & Fully automates visual analysis and anomaly detection with minimal user input. \\
    & & RM  & \fullcirc & The research field of visual AI security is still developing, with improvements in suspicious activity recognition. \\
    & & DP  & \fullcirc & Development potential includes task execution, sound analysis, and active security responses. \\
    & & IS  & \emptycirc & Improves the detection of physical attacks detection but does not consider network or software security. \\
    \cline{1-5}
    \multirow{6}{*}{9} &
    \multirow{6}{*}{~\citet{saha2023llm}}
      & ETD & \emptycirc & Does not detect or prevent external threats as it only critiques hardware design. \\
    & & IAD & \halfcirc & Detects internal vulnerabilities but is limited to hardware design flaws, not software or system anomalies. \\
    & & RA  & \halfcirc & Provides critiques based on user-defined security criteria. \\
    & & RM  & \fullcirc & Research in hardware security utilizing LLMs is still at an early stage. \\
    & & DP  & \fullcirc & Development potential includes automating design critiques and adapting diverse security standards. \\
    & & IS  & \fullcirc & Enhances hardware security and mitigates side-channel attacks and zero-day vulnerabilities. \\
    \cline{1-5}
    \multirow{6}{*}{10} &
    \multirow{6}{*}{ \makecell[c]{BERTIDS\\\citet{lira2024}}}
      & ETD & \halfcirc & Detects external threats through network log analysis but does not analyze live threats. \\
    & & IAD & \emptycirc & Does not detect internal anomalies as it only analyzes network logs. \\
    & & RA  & \fullcirc & Fully automates attack classification based on network log data with minimal user input. \\
    & & RM  & \emptycirc & The field of intrusion detection is well-established, leaving little room for novel research contributions. \\
    & & DP  & \halfcirc & Development potential for live network monitoring and automated task execution. \\
    & & IS  & \halfcirc & Limited impact on security as it does not offer new advancements beyond existing IDS. \\
    \cline{1-5}
    \multirow{6}{*}{11} &
    \multirow{6}{*}{~\citet{guastalla2023application}}
      & ETD & \halfcirc & Detects external threats (i.e., DDoS attacks) but does not extend to other types of attacks. \\
    & & IAD & \emptycirc & Does not detect internal anomalies. \\
    & & RA  & \fullcirc & Fully automates the detection of DDoS attacks with a minimum human involvement. \\
    & & RM  & \emptycirc & DDoS attack detection is a well-established research field, limiting its novelty. \\
    & & DP  & \halfcirc & Development potential to extend detection of other types of attacks. \\
    & & IS  & \fullcirc & An early example of LLM-based anomaly detection for IoT security. \\
    \hline
    \end{tabular}
    \end{threeparttable}
\end{table}
\end{landscape}
\begin{landscape}
\begin{table}
    \centering    
    \catcode`\_=12
    \catcode`\#=12
    \begin{threeparttable}
    \begin{tabular}{|r|c|c|c|l|}
    \hline    
    \multirow{6}{*}{12} &
    \multirow{6}{*}{\makecell[c]{SecurityBERT\\~\citet{ferrag2024revolutionizing}}}
      & ETD & \halfcirc & Detects external threats, such as malware and injection attacks, but does not automatically prevent them. \\
    & & IAD & \emptycirc & It only processes network logs and lacks system vulnerability analysis. \\
    & & RA  & \fullcirc & Automatically classifies external threats using network data, without human intervention. \\
    & & RM  & \emptycirc & Intrusion detection is a well-researched field with limited room for novel contributions. \\
    & & DP  &  \halfcirc & Development potential to integrate automated response mechanisms. \\
    & & IS  & \halfcirc & Limited impact as it lacks additional functionality beyond detection. \\
    \cline{1-5}
    \multirow{6}{*}{13} &
    \multirow{6}{*}{\makecell[c]{IDS-Agent\\\citet{li2024idsagent}}}
      & ETD & \fullcirc & Detects external threats by analyzing network traffic for suspicious patterns. \\
    & & IAD & \emptycirc & Focuses on monitoring network behaviors, not internal software vulnerabilities. \\
    & & RA  & \fullcirc & Fully automates intrusion detection with minimal human interaction. \\
    & & RM  & \fullcirc & Intrusion detection remains an active field of research, with evolving methods for network security. \\
    & & DP  & \fullcirc & Development potential for real-time detection and response automation. \\
    & & IS  & \fullcirc & Impactful by extending real-time network threat detection. \\
    \cline{1-5}
    \multirow{6}{*}{14} &
    \multirow{6}{*}{~\citet{islam2024llmpowered}}
      & ETD & \emptycirc & Focuses solely on internal vulnerability patching. \\
    & & IAD & \fullcirc & Detect and patch internal code vulnerabilities to prevent exploitation. \\
    & & RA  & \fullcirc & Fully automates detecting and patching vulnerabilities with minimal user input. \\
    & & RM  & \halfcirc & Vulnerability patching using LLMs is an emerging research field. \\
    & & DP  & \fullcirc & Optimize patching efficiency and integration with security frameworks. \\
    & & IS  & \fullcirc & Enhances efficiency in automatically identifying and patching vulnerabilities. \\
    \cline{1-5}
    \multirow{6}{*}{15} &
    \multirow{6}{*}{\makecell[c]{DefectHunter\\\citet{wang2023defecthunter}}}
      & ETD & \emptycirc & Strictly focuses on internal software security. \\
    & & IAD & \fullcirc & Detects and patches internal code vulnerabilities, preventing internal anomalies. \\
    & & RA  & \fullcirc & Fully automates vulnerability detection and patching, making it more efficient. \\
    & & RM  & \halfcirc & Secure coding is well-established while automated patching using LLMs is an emerging research area. \\
    & & DP  & \fullcirc & Development potential in integrating real-time security monitoring features. \\
    & & IS  & \fullcirc & Significant security impact as automated patching enhances efficiency and reduces vulnerabilities. \\
    \cline{1-5}
    \multirow{6}{*}{16} &
    \multirow{6}{*}{\makecell[c]{AttackGen\\\citet{Adams_2024}}}
      & ETD & \fullcirc & Provides external threat prevention through incident reports and mitigation strategies. \\
    & & IAD & \emptycirc & Does not detect internal anomalies. \\
    & & RA  & \fullcirc & Fully automates generating incident reports and playbooks with minimal user input. \\
    & & RM  & \fullcirc & Automated incident report generation is a new field enabled by LLMs. \\
    & & DP  & \fullcirc & High development potential, especially in integrating autonomous task execution. \\
    & & IS  & \fullcirc & Pioneers the use of LLMs for cyber security automation and report generation. \\    
    \cline{1-5}
    \multirow{6}{*}{17} &
    \multirow{6}{*}{~\citet{mcintosh2023harnessing}}
      & ETD & \fullcirc & Generates security policies to improve system security. \\
    & & IAD & \emptycirc & Focuses on policy generation rather than system vulnerability analysis. \\
    & & RA  & \fullcirc & Fully automates policy generation based on user queries, with no further input required. \\
    & & RM  & \fullcirc & LLM-driven cyber security policy generation is still emerging, making it a promising area of research. \\
    & & DP  & \halfcirc & Development potential is constrained, as its focuses strictly on policy creation without automated execution. \\
    & & IS  & \fullcirc & Improves efficiency and accuracy in cyber security policy development, streamlining industry practices. \\
    \hline
    \end{tabular}    
    \end{threeparttable}
\end{table}
\end{landscape}
\begin{landscape}
\begin{table}
    \centering    
    \catcode`\_=12
    \catcode`\#=12
    \begin{threeparttable}
    \begin{tabular}{|r|c|c|c|l|}
    \hline    
    \multirow{6}{*}{18} &
    \multirow{6}{*}{~\citet{bethany2024large}}
      & ETD & \halfcirc & Helps prevent external threats by generating spear-phishing emails for training. \\
    & & IAD & \emptycirc & Does not detect internal anomalies. \\
    & & RA  & \fullcirc & Fully automates phishing email generation with minimal human input. \\
    & & RM  & \emptycirc & Phishing is a well-researched field, and future developments shall focus more on efficiency. \\
    & & DP  & \halfcirc & Development potential includes refining content generation. \\
    & & IS  & \halfcirc & Limited impact since phishing is well-known, and awareness training already exists. \\
    \cline{1-5}
    \multirow{6}{*}{19} &
    \multirow{6}{*}{~\citet{yamin2024}}
      & ETD & \halfcirc & Prevents external threats by generating cyber security training scenarios. \\
    & & IAD & \emptycirc & Does not detect internal anomalies, focusing on training rather than identifying system vulnerabilities. \\
    & & RA  & \fullcirc & Fully automates scenario generation with minimal human input. \\
    & & RM  & \halfcirc & Cyber security training is an established field, but research on standardized best practices is ongoing. \\
    & & DP  & \emptycirc & Limited development potential in refining existing features rather than add new functionalities. \\
    & & IS  & \halfcirc & Limited to cyber security training. \\
    \cline{1-5}
    \multirow{6}{*}{20} &
    \multirow{6}{*}{\makecell[c]{LLM4Vuln\\\citet{sun2024llm4vuln}}}
      & ETD & \emptycirc & Does not detect or prevent external threats. \\
    & & IAD & \fullcirc & Detects internal anomalies by identifying vulnerabilities through automated analysis. \\
    & & RA  & \fullcirc & Fully automates vulnerability detection with minimal user intervention. \\
    & & RM  & \fullcirc & Vulnerability exploration is a continuously evolving research field. \\
    & & DP  & \fullcirc & Development potential to integrate automated patching and contextualized vulnerability remediation. \\
    & & IS  & \fullcirc & Automated vulnerability detection improves cyber security efficiency and effectiveness. \\
    \cline{1-5}
    \multirow{6}{*}{21} &
    \multirow{6}{*}{\makecell[c]{AutoAttacker\\\citet{xu2024autoattacker}}}
      & ETD & \fullcirc & Autonomously launches attacks to uncover known vulnerabilities in systems. \\
    & & IAD & \halfcirc & Exploits internal vulnerabilities but does not generate new attack methods. \\
    & & RA  & \fullcirc & Fully automates penetration testing by executing complex tasks and launching attacks. \\
    & & RM  & \fullcirc & Automated attack simulation is continuously evolving, offering new research directions. \\
    & & DP  & \fullcirc & High development potential to discover new vulnerabilities beyond known exploits. \\
    & & IS  & \halfcirc & Limited to commonly known vulnerabilities. \\
    \cline{1-5}
    \multirow{6}{*}{22} &
    \multirow{6}{*}{~\citet{tóth2024llms}}
      & ETD & \halfcirc & Detects external threats such as XSS and SQL injection but is limited to web-based applications. \\
    & & IAD & \halfcirc & Detects vulnerabilities in web applications. \\
    & & RA  & \fullcirc & Automates vulnerability detection with minimal human input, analyzing code for security flaws. \\
    & & RM  & \emptycirc & Web security is a mature field with existing best practices, limiting the novelty of this work. \\
    & & DP  & \fullcirc & High development potential to extend beyond Web security and enhance automation. \\
    & & IS  & \halfcirc & Limited impact as it primarily automates vulnerability detection. \\
    \cline{1-5}
    \multirow{6}{*}{23} &
    \multirow{6}{*}{~\citet{oliinyk2024fuzzing}}
      & ETD & \halfcirc & Prevents external threats by exposing vulnerabilities through fuzzing. \\
    & & IAD & \emptycirc & Only tests inputs that may cause crashes. \\
    & & RA  & \fullcirc & Fully automates the fuzzing process, executing crash tests with minimal human input. \\
    & & RM  & \halfcirc & While fuzzing is well-established, new opportunities emerge with recent advancement in GenAI and LLMs. \\
    & & DP  & \halfcirc & Limited development potential due to its focus on embedded Linux. \\
    & & IS  & \halfcirc & Increases fuzzing efficiency but does not provide solutions for discovered vulnerabilities. \\
    \hline
    \end{tabular}    
    \end{threeparttable}
\end{table}
\end{landscape}
\begin{landscape}
\begin{table}
    \centering    
    \catcode`\_=12
    \catcode`\#=12
    \begin{threeparttable}
    \begin{tabular}{|r|c|c|c|l|}
    \hline    
    \multirow{6}{*}{24} &
    \multirow{6}{*}{~\citet{happe2024llms}}
      & ETD & \halfcirc & Identifies and mitigates external threats effectively through penetration testing. \\
    & & IAD & \emptycirc & Does not detect internal software vulnerabilities or system anomalies. \\
    & & RA  & \halfcirc & Fully automates security testing and attack simulations with minimal human intervention. \\
    & & RM  & \emptycirc & Automated security testing is an evolving field with significant ongoing research. \\
    & & DP  & \fullcirc & High development potential to refine detection methods and extend capabilities. \\
    & & IS  & \emptycirc & Focused on penetration testing, but does not extend to broader security solutions. \\
    \cline{1-5}
    \multirow{6}{*}{25} &
    \multirow{6}{*}{\makecell[c]{LLMIF\\\citet{wangfuzzing2024}}}
      & ETD & \fullcirc & Detects external threats by exposing vulnerabilities in protocol implementation through fuzzing. \\
    & & IAD & \emptycirc & Does not analyze internal anomalies, as it is limited to network fuzzing. \\
    & & RA  & \fullcirc & Fully automates network protocol fuzzing with minimal user input. \\
    & & RM  & \halfcirc & Fuzzing research is still ongoing, but LLM-driven fuzzing remains relatively new. \\
    & & DP  & \halfcirc & Can be expanded with additional fuzzing techniques and broader protocol support. \\
    & & IS  & \halfcirc & Demonstrates the feasibility and effectiveness of automated protocol fuzzing. \\    
    \cline{1-5}
    \multirow{6}{*}{26} &
    \multirow{6}{*}{\makecell[c]{ChatAFL\\\citet{meng2024large}}}
      & ETD & \fullcirc & Effectively detects protocol-based security flaws through automated fuzzing. \\
    & & IAD & \emptycirc & Does not focus on internal vulnerabilities beyond protocol-level issues. \\
    & & RA  & \fullcirc & Fully automates protocol fuzzing, generating structured input cases independently. \\
    & & RM  & \fullcirc & The niche field is still relatively young, due to its recent emergence. \\
    & & DP  & \fullcirc & High potential for expanding into diverse network protocols and refining attack strategies. \\    
    & & IS  & \fullcirc & Enables stealth testing via mobile devices, inspiring further research in offensive and defensive security. \\    
    \cline{1-5}
    \multirow{6}{*}{27} &
    \multirow{6}{*}{\makecell[c]{FIAL\\\citet{androidfuzz2024}}}
      & ETD & \halfcirc & Detects external threats through network fuzzing but limited to network attack vector. \\
    & & IAD & \emptycirc & Does not detect internal threats or expose vulnerabilities in the system architecture or internal components. \\
    & & RA  & \fullcirc & Fully automates network fuzzing with minimal human intervention through an Android device. \\
    & & RM  & \halfcirc & Fuzzing is a well-established research field, but ongoing research using LLMs to enhance its automation. \\
    & & DP  & \fullcirc & High development potential to extend execution methods beyond Android (e.g., iOS and other devices). \\
    & & IS  & \fullcirc & Automates penetration testing but does not directly strengthen defensive measures. \\
    \cline{1-5}
    \multirow{6}{*}{28} &
    \multirow{6}{*}{~\citet{fang2024llm}}
      & ETD & \fullcirc & Detects and prevents external threats by automating attack simulations and discovering vulnerabilities. \\
    & & IAD & \emptycirc & Does not analyze internal configurations or detect internal anomalies. \\
    & & RA  & \fullcirc & Fully automates attack execution, launching exploits with minimal human input. \\
    & & RM  & \halfcirc & Penetration testing is a well-established research field, but LLM-based automation introduces new research opportunities. \\
    & & DP  & \halfcirc & Development potential includes its ability to discover vulnerabilities without CVEs. \\
    & & IS  & \halfcirc & Expands fuzzing capabilities but does not directly protect systems. \\
    \cline{1-5}
    \multirow{6}{*}{29} &
    \multirow{6}{*}{\makecell[c]{mGPTFuzz\\\citet{Maetal2024}}}
      & ETD & \emptycirc & Discovers vulnerabilities through fuzzing. \\
    & & IAD & \fullcirc & Highly effective at detecting internal vulnerabilities. \\
    & & RA  & \fullcirc & Fully automated with minimal human intervention. \\
    & & RM  & \emptycirc & Fuzzing is a well-established method with limited new research potential. \\
    & & DP  & \halfcirc & Development potential to extend beyond Matter to other protocols. \\
    & & IS  & \halfcirc & Remains focused on fuzzing without broader security applications. \\
    \hline
    \end{tabular}
    \end{threeparttable}
\end{table}
\end{landscape}
\begin{landscape}
\begin{table}
    \centering    
    \catcode`\_=12
    \catcode`\#=12
    \begin{threeparttable}
    \begin{tabular}{|r|c|c|c|l|}
    \hline    
    \multirow{6}{*}{30} &
    \multirow{6}{*}{\makecell[c]{PentestGPT\\\citet{deng2023pentestgpt}}}
      & ETD & \fullcirc & Detects and prevents external threats by simulating attacks, achieving high effectiveness. \\
    & & IAD & \halfcirc & Exploits internal vulnerabilities but does not have the ability to patch or fix them. \\
    & & RA  & \fullcirc & Fully automates penetration testing, requiring minimal human intervention. \\
    & & RM  & \fullcirc & Penetration testing remains an evolving research field, with continuous improvements in defense mechanisms. \\
    & & DP  & \halfcirc & Development potential to extend its capabilities to specialized areas such as IoT or ICS. \\
    & & IS  & \fullcirc & Improves penetration testing efficiency and strengthens system defense. \\
    \cline{1-5}
    \multirow{6}{*}{31} &
    \multirow{6}{*}{\makecell[c]{Net-GPT\\\citet{10419242}}}
      & ETD & \fullcirc & Detects external threats through man-in-the-middle (MitM) attacks to uncover system vulnerabilities. \\
    & & IAD & \emptycirc & Only focuses on network-based vulnerabilities. \\
    & & RA  & \fullcirc & Fully automates the generation and execution of network packet Mimicry for MitM attacks. \\
    & & RM  & \halfcirc & Using LLMs for MitM introduces new exploration opportunities. \\
    & & DP  & \fullcirc & High development potential to extend beyond UAV drone networks to other systems, including IoT. \\
    & & IS  & \halfcirc & Limited to exposing MitM vulnerabilities. \\
    \cline{1-5}
    \multirow{6}{*}{32} &
    \multirow{6}{*}{~\citet{Happe_2023}}
      & ETD & \fullcirc & Identifies and mitigates external threats effectively through penetration testing. \\
    & & IAD & \emptycirc & Does not detect internal software vulnerabilities or system anomalies. \\
    & & RA  & \fullcirc & Fully automates security testing and attack simulations with minimal human intervention. \\
    & & RM  & \fullcirc & Automated security testing is an evolving field with significant ongoing research. \\
    & & DP  & \fullcirc & High development potential to refine detection methods and extend capabilities. \\
    & & IS  & \halfcirc & Limited to forensic testing but does not proactively prevent new attacks. \\
    \cline{1-5}
    \multirow{6}{*}{33} &
    \multirow{6}{*}{~\citet{yang2023iot}}
      & ETD & \emptycirc & Does not exploit vulnerabilities through network or external attack vectors. \\
    & & IAD & \halfcirc & Detects internal threats using static code analysis. \\
    & & RA  & \fullcirc & Autonomously constrains vulnerabilities with minimal user interaction, achieving high detection accuracy. \\
    & & RM  & \emptycirc & Static code analysis is a mature research field, and this tool does not introduce novel research directions. \\
    & & DP  & \fullcirc & Development potential to automate vulnerability fixing and improve detection efficiency. \\
    & & IS  & \halfcirc & Limited to improving efficiency in static code analysis but does not contribute to broader security solutions. \\    
    \hline
    \end{tabular}    
    \begin{tablenotes}[para]        
        ETD: External Threat Detection.
        IAD: Internal Anomaly Detection.
        RA: Response Automation.
        RM: Research Maturity.
        DP: Development Potential.
        IS: Implact on Security.        
    \end{tablenotes}    
    \end{threeparttable}
\end{table}
\end{landscape}
\onecolumn
\section{Example Responses/Outputs for AttackGen Case Study}
Figure~\ref{fig:attackgen-default} provides the incident response plan generated with AttackGen~\citet{Adams_2024} without any modifications whereas Figure~\ref{fig:attackgen-iot} shows the incident response plan with IoT-specific modifications.
\begin{figure}[H]
\centering
\begin{tikzpicture}
[node distance=0.1cm, box/.style={rectangle, rounded corners, fill=blue!5, text width=\textwidth, minimum height=1cm, align=left}, title/.style={font=\large\bfseries}]
\node[box, fill=green!5] (DefaultAG){
\tiny
\noindent \textbf{Incident Response Testing Scenario: Dragonfly Supply Chain Compromise}

\smallskip
\noindent \textbf{Overview}

\noindent This incident response testing scenario is designed to evaluate the company’s incident and response capabilities against a sophisticated cyber attack by the threat actor group Dragonfly.
The attack vector is a supply chain compromise, which is a common tactic employed by Dragonfly to gain initial access to their targets.

\smallskip
\noindent \textbf{Objectives}
\begin{enumerate}[nosep]
    \item Assess the company’s ability to detect and respond to a supply chain compromise.
    \item Evaluate the effectiveness of communication and coordination among incident response team members.
    \item Test the procedures in place for mitigating and recovering from a supply chain compromise.
    \item Identify areas for improvement in the company’s incident response plan.
\end{enumerate}
\smallskip
\noindent \textbf{Scenario Details}

\hspace*{2mm} \textbf{Company Profile}
\begin{itemize}[nosep]
    \item Industry: Energy/Utilities
    \item Size: Medium (51-200 employees)
\end{itemize}
\hspace*{2mm} \textbf{Threat Actor Group}
\begin{itemize}[nosep]
    \item Name: Dragonfly
    \item Tactics, Techniques, and Procedures (TTPs): Utilizing supply chain compromise to gain initial access
\end{itemize}
\hspace*{2mm} \textbf{Kill Chain Phase}
\begin{itemize}[nosep]
    \item Initial Access: Supply Chain Compromise (T0862)   
\end{itemize}
\smallskip
\noindent \textbf{Scenario Timeline}

\hspace*{2mm} \textbf{Phase 1: Preparation}

\hspace*{2mm} Objectives:
\begin{itemize}[nosep]
    \item Ensure all incident response team members are aware of their roles and responsibilities.
    \item Verify that all necessary tools and resources are available and operational.
\end{itemize}

\hspace*{2mm} Actions:
\begin{enumerate}[nosep]
    \item Conduct a briefing session with the incident response team.
    \item Review the company’s incident response plan, focusing on supply chain compromises.
    \item Check the availability of forensic tools, communication channels, and backup systems.
\end{enumerate}

\hspace*{2mm} \textbf{Phase 2: Detection}

\hspace*{2mm} Objectives:
\begin{itemize}[nosep]
    \item Test the company’s ability to detect anomalies activities associated with a supply chain compromise.
\end{itemize}

\hspace*{2mm} Simulated Event:
\begin{itemize}[nosep]
    \item A third-party vendor supplying software to the company is compromised. The software update sent to the company contains a malicious payload.
\end{itemize}

\hspace*{2mm} Actions:
\begin{enumerate}[nosep]
    \item Deploy the compromised software update to a test environment.
    \item Monitor network traffic and system logs for signs of the compromise.
    \item Use indicators of compromise (IoCs) such as unusual network connections, unexpected file changes, and system behavior anomalies.
\end{enumerate}

\hspace*{2mm} \textbf{Phase 3: Containment}

\hspace*{2mm} Objectives:
\begin{itemize}[nosep]
    \item Evaluate the company’s capability to contain the threat and prevent further damage.
\end{itemize}

\hspace*{2mm} Simulated Event:
\begin{itemize}[nosep]
    \item The malicious payload attempts to establish a connection to an external command and control (C2) server.
\end{itemize}

\hspace*{2mm} Actions:
\begin{enumerate}[nosep]
    \item Isolate the affected systems from the network.
    \item Block outgoing connections to the identified C2 server.
    \item Notify relevant stakeholders of the incident.
\end{enumerate}

\hspace*{2mm} \textbf{Phase 4: Eradication}

\hspace*{2mm} Objectives:
\begin{itemize}[nosep]
    \item Test the company’s ability to remove the threat from affected systems.
\end{itemize}

\hspace*{2mm} Actions:
\begin{enumerate}[nosep]
    \item Perform a thorough analysis of the compromised systems to identify all malicious components.
    \item Remove the malicious payload and any associated malware.
    \item Apply patches and updates to close the vulnerability exploited by the attacker.
\end{enumerate}

\hspace*{2mm} \textbf{Phase 5: Recovery}

\hspace*{2mm} Objectives:
\begin{itemize}[nosep]
    \item Assess the company’s procedures for restoring normal operations.
\end{itemize}

\hspace*{2mm} Actions:
\begin{enumerate}[nosep]
    \item Restore affected systems from clean backups.
    \item Monitor the systems to ensure that no residual threats remain.
    \item Gradually reconnect the systems to the network.
\end{enumerate}

\hspace*{2mm} \textbf{Phase 6: Lessons Learned}

\hspace*{2mm} Objectives:
\begin{itemize}[nosep]
    \item Review the incident response process to identify strengths and areas for improvement.
\end{itemize}

\hspace*{2mm} Actions:
\begin{enumerate}[nosep]
    \item Conduct a debriefing session with all incident response team members.
    \item Document the incident and response actions taken.
    \item Update the incident response plan based on the lessons learned.
\end{enumerate}

\smallskip
\noindent \textbf{Evaluation Criteria}

\begin{enumerate}[nosep]
    \item Detection:
        \begin{itemize}[nosep]
        \item Time taken to detect the compromise.
        \item Accuracy in identifying IoCs.
        \end{itemize}
    \item Containment:
        \begin{itemize}[nosep]
        \item Speed and effectiveness of containment measures.
        \item Communication and coordination among team members.
        \end{itemize}
    \item Eradication:
        \begin{itemize}[nosep]
        \item Thoroughness in removing all malicious components.
        \item Effectiveness of patching and updating procedures.
        \end{itemize}
    \item Recovery:
        \begin{itemize}[nosep]
        \item Efficiency in restoring normal operations.
        \item Monitoring for residual threats.
        \end{itemize}
    \item Lessons Learned:
        \begin{itemize}[nosep]
        \item Quality of the post-incident analysis.
        \item Implementation of improvements to the incident response plan.
        \end{itemize}
\end{enumerate}

\smallskip
\noindent \textbf{Conclusion}

This incident response testing scenario provides a comprehensive framework to evaluate the company’s readiness against a supply chain compromise by Dragonfly.
By following the outlined phases and actions, the company can identify strengths and areas for improvement in their incident response capabilities, ensuring they are better prepared for future threats.
};

\end{tikzpicture}
    \caption{Incident Response Testing Scenario for Dragonfly Supply Chain Compromise using AttackGen without Modifications}
    \label{fig:attackgen-default}
\end{figure}
\onecolumn
\begin{figure}[H]
\centering
\begin{tikzpicture}
[node distance=0.1cm, box/.style={rectangle, rounded corners, fill=blue!5, text width=\textwidth, minimum height=1cm, align=left}, title/.style={font=\large\bfseries}]
\node[box] (IoTAG){
\tiny
\noindent \textbf{Incident Response Testing Scenario: Dragonfly Supply Chain Compromise with IoT-Powered PLC}

\smallskip
\noindent \textbf{Overview}

\noindent This incident response testing scenario is designed to evaluate the company’s readiness and response capabilities against a sophisticated cyber attack by the threat actor group Dragonfly. This attack vector is a supply chain compromise, which is a common tactic employed by Dragonfly to gain initial access to their targets. The scenario will specifically focus on the compromise of an IoT-based Programmable Logic Controller (PLC) that is critical to the company’s operations and is connected to the internet.

\smallskip
\textbf{Objectives}
\begin{enumerate}[nosep]
    \item Assess the company’s ability to detect and respond to a supply chain compromise involving an IoT-powered PLC.
    \item Evaluate the effectiveness of communication and coordination among incident response team members.
    \item Test the procedures for monitoring and recovering from a supply chain compromise affecting critical IoT devices.
    \item Identify areas for improvement in the company’s incident response plan.
\end{enumerate}

\smallskip
\textbf{Scenario Details}

\hspace*{2mm} \textbf{Company Profile}
\begin{itemize}[nosep]
    \item Industry: Energy / Utilities 
    \item Size: Medium (51-200 employees) Critical Asset: IoT powered PLC connected to the internet
\end{itemize}
\hspace*{2mm} \textbf{Threat Actor Group}
\begin{itemize}[nosep]
    \item Name: Dragonfly
    \item Tactics, Techniques, and Procedures (TTPs): Utilizing supply chain compromise to gain initial access
\end{itemize}
\hspace*{2mm} \textbf{Kill Chain Phase}
\begin{itemize}[nosep]
    \item Initial Access: Supply Chain Compromise (T0862)  
\end{itemize}
\smallskip
\textbf{Scenario Timeline}

\hspace*{2mm} \textbf{Phase 1: Preparation}

\hspace*{2mm} Objectives:
\begin{itemize}[nosep]
    \item Ensure all incident response team members are aware of their roles and responsibilities.
    \item Verify that all necessary tools and resources are available and operational.
\end{itemize}

\hspace*{2mm} Actions:
\begin{enumerate}[nosep]
    \item Conduct a briefing session with the incident response team.
    \item Review the company’s incident response plan, focusing on supply chain compromise scenarios, particularly those involving IoT devices.
    \item Check the availability of forensic tools, communication channels, and backup systems. Ensure proper monitoring mechanisms are in place for IoT devices, especially the PLC.
\end{enumerate}

\hspace*{2mm} \textbf{Phase 2: Detection}

\hspace*{2mm} Objectives:
\begin{itemize}[nosep]
    \item Test the company’s ability to detect anomalous activities associated with a supply chain compromise of an IoT device.
\end{itemize}

\hspace*{2mm} Simulated Event:
\begin{itemize}[nosep]
    \item A third-party vendor supplying firmware to the IoT-powered PLC is compromised. The firmware update sent to the company contains a malicious payload.
\end{itemize}

\hspace*{2mm} Actions:
\begin{enumerate}[nosep]
    \item Deploy the compromised firmware update to a test environment with an IoT-powered PLC.
    \item Monitor network traffic, system logs, and PLC activity for signs of the compromise.
    \item Identify indicators of compromise (IoCs) such as unusual network connections, unexpected firmware changes, and anomalous PLC behavior.
\end{enumerate}

\hspace*{2mm} \textbf{Phase 3: Containment}

\hspace*{2mm} Objectives:
\begin{itemize}[nosep]
    \item Evaluate the company’s capability to contain the threat and prevent further damage, particularly to critical IoT devices.
\end{itemize}

\hspace*{2mm} Simulated Event:
\begin{itemize}[nosep]
    \item The malicious payload attempts to establish a connection to an external command and control (C2) server and manipulate the PLC’s operations.
\end{itemize}

\hspace*{2mm} Actions:
\begin{enumerate}[nosep]
    \item Isolate the affected PLC and any other compromised systems from the network.
    \item Block outgoing connections to the identified C2 server.
    \item Notify relevant stakeholders of the incident, emphasizing the critical nature of the compromised IoT device.
\end{enumerate}

\hspace*{2mm} \textbf{Phase 4: Eradication}

\hspace*{2mm} Objectives:
\begin{itemize}[nosep]
    \item Test the company’s ability to remove the threat from affected systems, including the critical IoT powered PLC.
\end{itemize}

\hspace*{2mm} Actions:
\begin{enumerate}[nosep]
    \item Perform a thorough analysis of the compromised PLC and other affected systems to identify all malicious components.
    \item Remove the malicious payload and any associated malware from the PLC.
    \item Apply patches and updates to close the vulnerability exploited by the attacker, ensuring the integrity of the firmware on the PLC.
\end{enumerate}

\hspace*{2mm} \textbf{Phase 5: Recovery}

\hspace*{2mm} Objectives:
\begin{itemize}[nosep]
    \item Assess the company’s procedures for restoring normal operations, focusing on the critical IoT-powered PLC.
\end{itemize}

\hspace*{2mm} Actions:
\begin{enumerate}[nosep]
    \item Restore the PLC from a clean backup, ensuring its firmware is uncompromised.
    \item Monitor the PLC and other systems to ensure that no residual threats remain.
    \item Gradually reconnect the PLC and other affected systems to the network.
\end{enumerate}

\hspace*{2mm} \textbf{Phase 6: Lessons Learned}

\hspace*{2mm} Objectives:
\begin{itemize}[nosep]
    \item Review the incident response process to identify strengths and areas for improvement, with a focus on IoT devices.
\end{itemize}

\hspace*{2mm} Actions:
\begin{enumerate}[nosep]
    \item Conduct a debriefing session with all incident response team members, emphasizing the response to the compromised IoT-powered PLC.
    \item Document the incident and response actions taken, particularly those involving IoT devices.
    \item Update the incident response plan based on the lessons learned, including specific measures for protecting and responding to incidents involving IoT devices.
\end{enumerate}

\smallskip
\textbf{Evaluation Criteria}

\begin{enumerate}[nosep]
    \item Detection:
        \begin{itemize}[nosep]
        \item Time taken to detect the compromise.
        \item Accuracy in identifying IoCs, especially those related to IoT devices.
        \end{itemize}
    \item Containment:
        \begin{itemize}[nosep]
        \item Speed and effectiveness of containment measures.
        \item Communication and coordination among team members, with a focus on IoT device management.
        \end{itemize}
    \item Eradication:
        \begin{itemize}[nosep]
        \item Thoroughness in removing all malicious components.
        \item Effectiveness of patching and updating procedures, particularly for IoT devices.
        \end{itemize}
    \item Recovery:
        \begin{itemize}[nosep]
        \item Efficiency in restoring normal operations.
        \item Monitoring for residual threats, especially in IoT devices.
        \end{itemize}
    \item Lessons Learned:
        \begin{itemize}[nosep]
        \item Quality of the post-incident analysis.
        \item Implementation of improvements to the incident response plan, including IoT-specific considerations.
        \end{itemize}
\end{enumerate}

\smallskip
\textbf{Conclusion}

This updated incident response testing scenario provides a comprehensive framework to evaluate the company’s readiness against a supply chain compromise by Dragonfly, with a specific focus on a critical IoT-powered PLC.
By following the outlined phases and criteria, the company can identify strengths and areas for improvement in their incident response capabilities, ensuring they are better prepared for future threats.
};

\end{tikzpicture}
    \caption{Incident Response Testing Scenario for Dragonfly Supply Chain Compromise Using AttackGen with IoT-Specific Modifications}
    \label{fig:attackgen-iot}
\end{figure}

%





\end{document}